\documentclass[prd,twocolumn,showpacs,showkeys,preprintnumbers,floatfix,superscriptaddress,longbibliography]{revtex4-1}

\usepackage{amsfonts} 
\usepackage{amssymb} 
\usepackage{amsmath} 
\usepackage{graphicx} 
\usepackage{subfigure} 
\usepackage{array} 
\usepackage{dcolumn} 
\usepackage{bm} 
\usepackage{latexsym} 
\usepackage{longtable} 
\usepackage{hyperref} 
\usepackage{bbold}
\usepackage{mathtools}
\usepackage{listings}

\usepackage[utf8]{inputenc}
\usepackage[braket]{qcircuit}
\usepackage[dvipsnames,usenames]{xcolor}

\newcommand{\GF}{G_{\text{F}}}
\newcommand{\svac}{\sin(2\theta)}
\newcommand{\cvac}{\cos(2\theta)}
\newcommand{\tvac}{\theta}
\newcommand{\A}{\text{A}}
\newcommand{\B}{\text{B}}
\newcommand{\Pv}{\vec{P}}
\newcommand{\Bv}{\vec{B}}
\newcommand{\Pva}{\vec{P}_{\A}}
\newcommand{\Pvb}{\vec{P}_{\B}}
\newcommand{\ex}{\hat{e}_{1}}
\newcommand{\ey}{\hat{e}_{2}}
\newcommand{\ez}{\hat{e}_{3}}
\newcommand{\nh}{\hat{n}}

\newcommand{\vvec}{\mathbf{v}}
\newcommand{\I}{\text{i}}
\newcommand{\dd}{\text{d}}

\newcommand{\Jv}{\vec{J}}
\newcommand{\sigv}{\vec{\sigma}}
\usepackage[normalem]{ulem}

\newcommand{\IQUS}{{InQubator for Quantum Simulation (IQuS), Department of Physics, University of Washington, Seattle, WA 98195, USA}}
\newcommand{\UNITN}{{Dipartimento di Fisica, University of Trento, via Sommarive 14, I–38123, Povo, Trento, Italy}}
\newcommand{\TIFPA}{INFN-TIFPA Trento Institute of Fundamental Physics and Applications,  Trento, Italy}

\begin{document}
\title{Classical and Quantum Evolution in a Simple Coherent Neutrino Problem}
\author{Joshua D. Martin}
\affiliation{ Theoretical Division, Los Alamos National Lab, Los Alamos, NM, 87545}
\author{A. Roggero}
\affiliation{\IQUS}
\affiliation{\UNITN}
\affiliation{\TIFPA}
\author{Huaiyu Duan}
\affiliation{Department of Physics and Astronomy, University of New Mexico, Albuquerque, New Mexico 87131, USA}
\author{J. Carlson}
\affiliation{ Theoretical Division, Los Alamos National Lab, Los Alamos, NM, 87545}
\author{V. Cirigliano}
\affiliation{ Theoretical Division, Los Alamos National Lab, Los Alamos, NM, 87545}

\preprint{IQuS@UW-21-016}

\date{\today}

\begin{abstract}
The extraordinary neutrino flux produced in extreme astrophysical environments like the early universe, core-collapse supernovae and neutron star mergers may produce coherent quantum neutrino oscillations on macroscopic length scales. The Hamiltonian describing this evolution can be mapped into quantum spin models with all-to-all couplings arising from neutrino-neutrino forward scattering. To date many studies of these oscillations have been performed in a mean-field limit where the neutrinos time evolve in a product state.   

In this paper we examine a simple two-beam model evolving from an initial product state and compare the mean-field and many-body evolution.  The symmetries in this model allow us to solve the real-time evolution for the quantum many-body system for hundreds or thousands of spins, far beyond what would be possible in a more general case with an exponential number ($2^N$) of quantum states.  We compare mean-field and many-body solutions for different initial product states and ratios of one- and two-body couplings, and find that in all cases in the limit of infinite spins the mean-field (product state) and many-body solutions coincide for simple observables.  
This agreement can be understood as a consequence of the fact that the typical initial condition represents a very local but dense distribution about a mean energy in the spectrum of the Hamiltonian.
We explore quantum information measures like entanglement entropy and purity of the many-body solutions, finding intriguing relationships between the quantum information measures and the dynamical behavior of simple physical observables.
\end{abstract}

\maketitle

\section{Introduction}

In core collapse supernovae and binary neutron star mergers 
neutrinos are emitted in extremely high number densities, and they 
can have a nontrivial impact on the chemical and hydrodynamical 
evolution of these environments.  Emitted neutrinos can affect
the neutron-to-proton ratio thereby impacting nucleosynthesis 
processes, and they can likely transport energy and reheat the shock 
formed during a core collapse supernova explosion 
\cite{RevModPhys.62.801,Janka:2006fh,Woosley:2005,Hoffman:1997}.  In such dense 
neutrino gases, the neutrinos can experience coherent forward 
scattering with other local neutrinos and thus generate a 
self-coupled evolution in the flavor content of the gas 
\cite{Qian:1994wh,Qian:1995,Balantekin:2006tg}.  

Significant work has gone into the study of these dense neutrino gases 
under a variety of simplifying assumptions and imposed symmetries 
almost exclusively in the mean-field approximation, which
is equivalent to the time evolution of the system within the
space of product states of single-neutrino spinors. These dense neutrino 
systems exhibit a rich variety of phenomena in the evolution and 
transport of their flavor content.  Such phenomena include swaps 
between the initial spectra of different neutrino flavors 
\cite{Duan:2007mv,Raffelt:2007cb,Martin:2019dof}, collective 
and synchronized evolution \cite{Raffelt:2007xt,Hannestad:2006nj}, 
coherently transported flavor waves \cite{Martin:2019gxb, PhysRevD.104.103003}, 
spontaneous symmetry breaking \cite{Chakraborty:2015tfa, Yi:2019hrp}, and the generation of very fine scale 
spatial flavor structure analogous to fluid turbulence \cite{Chakraborty:2015tfa, Mirizzi:2015fva}.

Recent work has suggested however that in some cases behavior which is 
observed in systems analyzed using the mean-field approximation may 
deviate significantly from that seen in equivalent many-body solutions 
which retain all quantum correlations, requiring in general a basis size growing exponentially with system size \cite{Patwardhan2019,patwardhan2021spectral,Rrapaj2020}.

The dense neutrino Hamiltonian governing the neutrino flavor evolution (in the two flavor approximation) is equivalent to a Heisenberg-like spin model with 
long-range (in flavor space) neutrino-neutrino flavor exchange interactions and spatially varying single-particle ``magnetic" fields. The time evolution of these systems is of great
interest in the condensed matter, AMO, and quantum computing
communities (see e.g.~\cite{Zhang2017,xu2020}).  This Hamiltonian can be simply mapped
to qubits, though with the caveat that the two-body 
interactions are long-ranged. Each spin represents the neutrino
field at a specific magnitude and direction of momentum, with the spin degrees of freedom representing the neutrino flavor.

We assume an initial product state of single-neutrino spinors and study
the evolution after a rapid quench to the full Hamiltonian.
An initial product state is appropriate if one-body evolution dominates in the interior of the proto-neutron star
where coherent forward scattering on the dense background of charged leptons dominates.  The rapid quench is
clearly an approximation but is very useful for testing
the resulting dynamics. In particular we are interested in
studying the evolution of quantum information measures in this
system, including entanglement entropy and purity.  We also study simple one-body observables including
the time-dependent flavor expectation value as a function
of momentum direction and energy. The relation between the full quantum dynamics and the product state evolution can be examined using these observables and the 
quantum information content of the evolved states. Gate-based quantum computers
can in principle solve for the many-body dynamics, and 
emulators including trapped-ion and Rydberg cold atom systems
should be able to help investigate closely related problems in quantum spin dynamics~\cite{Monroe2021,Ebadi2021}.

In this work we investigate the simplest dense neutrino model
which includes neutrino-neutrino coherent forward scattering. 
This model describes a dense gas of neutrinos with only two momentum directions and 
energies.  In Sec.~\ref{sec:models} we introduce the
many-body Hamiltonian and in Sec.~\ref{sec:moments} our simple initial condition.
The initial state consists of a product state of the two 
beams with fully aligned spins within each beam.   The 
initial-state and Hamiltonian symmetry with respect to interchange of neutrinos within each beam enables exact
solutions for large numbers of neutrinos since the number
of quantum amplitudes grows only polynomially with system
size rather than the exponential growth in the general case.

The initial state energy and higher moments of the Hamiltonian
play important roles in the full quantum evolution.
Employing our initial state, we calculate  the first four moments of the many-body Hamiltonian, 
and we demonstrate that in the 
limit of a large number of spins the initial product state has a Gaussian distribution 
in the energy eigenspace of the Hamiltonian.  We will use the scaling properties of the variance to 
construct a heuristic measure describing when mean-field-like behavior should  be expected to emerge in a many-body calculation. 

In Sec.~\ref{sec:dynamics} we compare the dynamical evolution of both the many-body and mean-field evolution. Using our proposed heuristic, we argue that the mean-field approximation correctly predicts the averages of simple one-body 
operator expectation values in the large many-body system limit except in the 
special case that the initial product state is an eigenstate of the one-body Hamiltonian.

In Sec.~\ref{sec:entanglement} we investigate the full time evolution of this system including two primary quantum information measures, the bipartite entanglement entropy of
the two beams and the purity of 
individual neutrino quantum states. 
Finally, in Sec.~\ref{sec:conclusions} we present our conclusions and thoughts regarding 
future work.

\section{Neutrino Hamiltonian and Two-Beam Geometry}
\label{sec:models}

The Hamiltonian which governs the evolution of the flavor content 
of the dense neutrino gas comes in three pieces.  First is due 
to the mismatch between the mass and weak flavor states. The second 
is generated from the coherent forward scattering of the neutrinos 
off of the local charged leptons, which can result in the famous MSW 
resonance.  Finally, through the weak interaction, the neutrinos can 
experience coherent forward scattering with other present neutrinos. 
In the rest of this work, we will work in the two flavor approximation, denoting 
the electron flavor neutrinos $\nu_{e}$ and the second ``$x$" flavor neutrinos as 
$\nu_{x}$.  We also note that this $x$ state should be understood 
to represent a linear combination of the physical muon and $\tau$ neutrino 
flavor states.

The Hamiltonian we will study has the form~\cite{Pehlivan2011}
\begin{equation}\label{eq:manyBodyH} 
    H = \sum_{i} \left[ \frac{\omega_{i}}{2} \Bv \cdot \sigv_{i} 
        \right]
        + \frac{\sqrt{2} \GF}{2 V} \sum_{i < j} 
            \left( 1 - \vvec_{i} \cdot \vvec_{j} \right)\sigv_{i} \cdot \sigv_{j} .
\end{equation}
Here, the sums are over all of the flavor spins and the vector operators $\sigv_i=(\sigma_i^x,\sigma_i^y,\sigma_i^z)$ are constructed with the usual Pauli matrices acting on the $i^{\text{th}}$ neutrino amplitude.  The vacuum oscillation frequencies are
$\omega_{i} = \Delta m^{2} / 2 E_{i} $ where $\Delta m^{2}$ is the mass squared splitting, and 
$E_{i}$ is the energy of the $i^{\text{th}}$ neutrino.  We choose to work in the mass basis, 
such that $\Bv = - \ez$.  $\GF$ is the usual Fermi coupling constant, and the term proportional to $\GF$ is generated 
by the neutral current coherent forward scattering between the neutrinos along different trajectories ($\vvec_i$). We have not included the usual 
electron coherent forward scattering term proportional to 
$\GF n_{e}$ (where $n_{e} = (n_{e^{-}} - n_{e^{+}})$ is the the net electron density).
We will mimic the inclusion of the term generated by a dense background of electrons by assuming that its effect 
is to reduce the effective vacuum mixing angle between the mass and flavor states. A direct inclusion of this term would be preferable since it is not possible in general to properly quantify the amount of this suppression. Due to the large magnitude of the matter term compared to the other contributions, a naive implementation of the full time-evolution incurs a considerable increase of the simulation cost. This problem can be circumvented using algorithms to perform simulations in the interaction picture (see e.g.~\cite{rajput2021hybridized}) and we plan to leverage this technology in future work. We assume 
that the presence of any other charged leptons is negligible.

Determining the flavor evolution of the dense neutrino gas, even under the assumptions of 
homogeneity and isotropy, is prohibitively difficult.  For an arbitrary initial 
condition describing the initial flavor states of $N$ neutrinos, the time evolution 
of $2^{N}$ complex amplitudes must be tracked consistently.  In the following work we will study 
the flavor dynamics of a system which is approximated as two ``beams'' of neutrinos.  
In this approximation, there are only two distinct velocities $\vvec_{\A}$ and 
$\vvec_{\B}$, so we can extract the geometric factor $1-\vvec_{i} \cdot \vvec_{j}$ 
from the neutrino-neutrino coherent forward scattering potential.  We also assume 
that within each beam there are monochromatic neutrinos such that we only retain two 
distinct vacuum oscillation frequencies, $\omega_{\A}$ and $\omega_{\B}$.

With the momentum geometry and energies specified, we will work 
in the frame which rotates about the $\Bv$ axis with frequency 
\begin{equation*}
    \frac{(\omega_{\A} + \omega_{\B})}{2} 
\end{equation*}
such that we drop the component of the vacuum oscillation 
Hamiltonian which is common to both beams.  The two body Hamiltonian is characterized by the strength 
\begin{equation}
    \mu = \frac{\sqrt{2} \GF N}{V} \left( 1 - \vvec_\A \cdot \vvec_\B \right) ,
\end{equation}
and we will measure all other energies and times in units of $\mu$.  We thus define 
$\Omega = (\omega_{\A} - \omega_{\B}) / \mu$ and express the two-beam Hamiltonian for the quantum many-body problem in 
units of $\mu$ as 
\begin{equation} \label{eq:twoBeamH}
    \frac{H}{\mu} = \frac{\Omega}{2} \Bv \cdot \left( \Jv_{\A} - \Jv_{\B} \right)
        + \frac{ 2 }{ N } \Jv_{\A} \cdot \Jv_{\B} .
\end{equation}
where $\Jv_{\A / \B} = \sum_{i \in \A / \B} \sigv_{i}/2$.
We note that the Hamiltonian in Eq.~\eqref{eq:twoBeamH} is integrable and a 
complete solution could, in principle, be obtained using the Bethe ansatz~\cite{Pehlivan2011,Birol:2018qhx,Patwardhan2019}.  Having normalized all energies to the 
characteristic scale of the neutrino-neutrino forward scattering term, we set $\mu = 1$ thereby suppressing explicit 
dependence on $\mu$ throughout the rest of this work.

\section{Initial Product States and Energy Moments}
\label{sec:moments}

The initial conditions we will study are product states of the individual spins with aligned spins within each beam.  This is a highly simplified case of a more realistic initial state in which, for example, oscillations are suppressed by the large matter density near the surface of a proto-neutron star, but the decoupling regime at the surface will be energy and flavor-dependent.  It has the advantage of making it easy to compare the evolution of the mean-field and many-body case starting from the same initial state.
The symmetries in this initial state can also be exploited to treat the many-body dynamics very efficiently.

We write our initial state as
\begin{equation} \label{eq:genIC}
    \ket{\Psi} = \ket{\nh_{\A}}^{\otimes N_{\A}}
        \ket{\nh_{\B}}^{\otimes N_{\B}} .
\end{equation}
The unit vectors $\nh_{\A / \B}$ are parameterized by azimuthal and 
polar angles $\theta_{\A / \B}$ and $\phi_{\A / \B}$, and the 
individual single particle states are written in terms of these angles 
as
\begin{equation} \label{eq:oneSpin}
    \ket{\nh_{\A / \B}} = \cos\left( \frac{\theta_{\A / \B}}{2}\right) \ket{\nu_{1}}
        +\sin\left( \frac{\theta_{\A / \B}}{2}\right) e^{\I \phi_{\A / \B}} \ket{\nu_{2}}
\end{equation}
where $\ket{\nu_{1}}$ and $\ket{\nu_{2}}$ are the mass eigenstates of the single 
neutrino vacuum Hamiltonian.

\begin{figure}
    \begin{center}
        \includegraphics[width=0.48\textwidth]{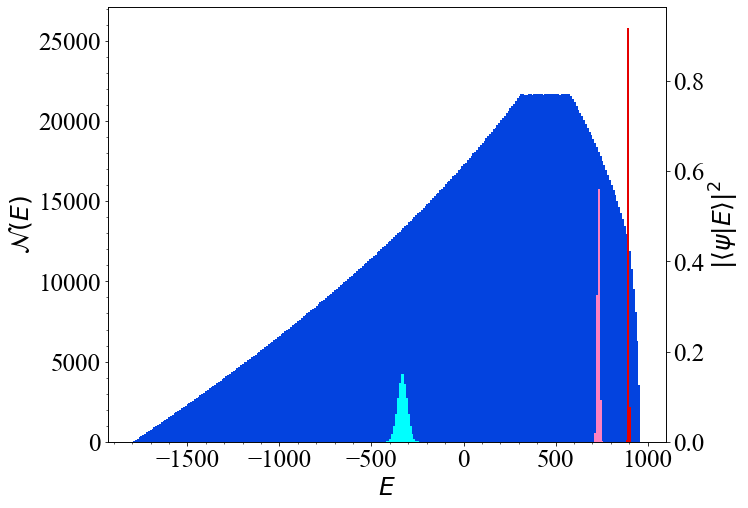} 
    \end{center}
    \caption{ Histogram of the number of energy states of the many-body Hamiltonian (blue) in the $J_{\A} = N_{\A}/2$, $J_{\B} = N_{\B}/2$ subspace for $N = 3600$ spins.  Energy bins have a width of 10 in units of 
    $\mu$.  The energy distribution corresponds to the choice of energy asymmetry ($\Omega$) and population fraction for case 2 the bipolar mode solution as specified in table \ref{tab:bipolar}. Also shown are three initial conditions projected over the energy spectrum (red, pink, and cyan histograms).  The red histogram corresponds to an initial polarization in energy space which results in bipolar oscillations in the large $N$ limit (case 2 in table \ref{tab:bipolar}).  Similarly, the pink corresponds to an initial polarization which results in collective precession of the flavor polarization vectors in the large $N$ limit (case 2 in table \ref{tab:precession}).  Finally, the cyan represents randomly chosen polarizations for the $\nh_{\A / \B}$ unit vectors (case 6 in table \ref{tab:random}).  
    }
    \label{fig:d_of_states}
\end{figure}

This initial condition is highly symmetric, and as such it accesses only a tiny fraction of the eigenstates of the 
total many body Hamiltonian. We observe that the number of energy states with nonzero overlap with this initial condition scales at most as $\sim N^{3/2}$ rather than exponentially in $N$, which we will justify in the following paragraphs. 
We will express the initial condition in the angular momentum basis $\ket{J_\A,M_{\A}}$ of each block of spins such that 
\begin{equation} \label{eq:genIC2}
    \ket{\Psi} = \sum_{M_{\A},M_{\B}} c_{M_{\A},M_{\B}} \ket{J_{\A},M_{\A}} \otimes \ket{J_{\B},M_{\B}} .
\end{equation}
We also see that the Hamiltonian keeps invariant the individual squared angular momentum of each block, $J_{\A/\B}^{2}$, 
and the total $\ez$ projection $J_{3} = M_{\A} + M_{\B}$ (i.e. the projection into $\Bv = -\ez$ in the mass basis) and that in this choice of basis the many-body 
Hamiltonian is tri-diagonal.  The initial condition is a state with maximal $J_\A^2$ and $J_\B^{2}$, so we therefore only need to determine with which total angular momentum projection 
$J_3$ subspaces our initial state has appreciable nonzero overlap, and we can then efficiently diagonalize those subspaces due to their tri-diagonal structure using the subroutine
\verb|eigh_tridiagonal| provided by SciPy \cite{2020SciPy-NMeth} (see also~\cite{xiong2021manybody}). 
Furthermore, from the conserved quantities of the Hamiltonian and the structure of the general form of our
initial condition (Eq.~\ref{eq:genIC2}) we observe that the dimensionality of the accessible Hilbert space 
scales at most as $N^{2}$
In fig. \ref{fig:d_of_states} we show the total distribution of energy eigenstates as a histogram for all possible $J_3$ subspaces of the 
Hamiltonian with $J_{\A / \B} = N_{\A / \B} / 2$.

In this subsection, we compute the energy distribution of the initial product state in terms of moments of the Hamiltonian calculated with respect to our initial state.
For our time-independent Hamiltonian, energy conservation plays an important role in 
the evolution of the system.  For specific initial states near the extremes of the spectra, phenomena such as dynamical phase transitions may be present~\cite{roggero2021dynamical}.  For this all-to-all Hamiltonian interaction, as we show below, the full spectrum has a range that is proportional to $N$ while the width of the energy distribution of an initial product state is proportional to $\sqrt{N}$ while the energy level spacing for a given total $J_3$ is approximately constant for large $N$. The energy level spacing summed over all $J_3$ is proportional to $1/N$. This behavior is also seen in a typical spin models with short-range interactions. 
In this subsection we discuss the moments of the two-beam model, but these can be easily computed for more general cases.

We will proceed by calculating the 
expectation values of the first two moments, and the third and fourth central moments, of the initial condition in the 
spectrum of the Hamiltonian.  The expectation
value of the Hamiltonian is:
\begin{align} \label{eq:expH}
    \langle H \rangle = \frac{N}{4} \Big[ 
        &\Omega \vec{B} \cdot \left( \nh_{\A} \frac{N_\A}{N} - \nh_{\B} \frac{N_\B}{N} \right) \nonumber \\
        &+ 2 \nh_{\A} \cdot \nh_{\B} \frac{N_\A N_\B}{N^{2}} \Big] .
\end{align} 

The expectation value of $\langle H^2 \rangle$
can be computed by expanding the terms in the square of the Hamiltonian, and the surviving terms in the variance arise
only from terms with repeated spin indices; for operator products applied on different spin components of the state the expectation value of the product is the same as the product of the expectation values. 
In general terms with more repeated spin indices will produce lower powers of $N$ in the $n^{\text{th}}$ central moment of the Hamiltonian.
The variance can be written in the form:
\begin{equation} \label{eq:varH}
    \Delta H^{2} = c_{1} N + c_{0} .
\end{equation}

The term $c_{0}$ has the form
\begin{align} \label{eq:c0}
    c_{0} = \frac{N_\A N_\B}{4N^{2}} \Big( 
            1 -  (\nh_\A \cdot \nh_\B)
        \Big)^2  .
\end{align}
which vanishes when $\nh_{\A} = \nh_{\B}$ since this state is an eigenstate of total spin.  We also note that it contains no term proportional to $\Omega$, therefore it 
stems only from the $\nu-\nu$ interaction term in the Hamiltonian.  As the one body term alone cannot generate inter-particle correlation effects, if $c_{0}$ dominates the variance for some finite value of $N$, we expect to be in the regime in which many-body effects will be significant due to the finite size.  It is therefore important to study the ratio $c_{0}/(c_{1} N)$ as this will control the size of $N$ where mean-field like behavior (which works directly in the $N\to\infty$  limit) can possibly emerge. 
Critically, if $c_{1}$ vanishes for some choice of parameters, we expect that there exists no value of $N$ such that the many-body and mean field solutions will agree.

Next we find that $c_{1}$ is a $2^{\text{nd}}$ order polynomial in $\Omega$. The second order term comes from the square of the one-body term, the zeroth order from the square of the two-body term, and the first order from the product of the two.  We write the variance as:
\begin{equation}
    c_{1} = c_{1,2} \Omega^{2} + c_{1,1} \Omega + c_{1,0} .
\end{equation}
 For
arbitrary initial polarizations we find that the coefficients are:
\begin{align}
    c_{1,2} &= \frac{1}{16} \Big[\frac{N_\A}{N} |\Bv \times \nh_{\A}|^{2}
        + \frac{N_\B}{N} |\Bv \times \nh_{\B}|^{2} \Big] , \\
    c_{1,1} &= \frac{N_\A N_\B}{4N^{2}} 
    \Bv \cdot (\nh_\B - \nh_\A) (1+\nh_\A \cdot \nh_\B) ,\\
    \text{and } c_{1,0} &= \frac{N_\A N_\B}{4N^{2}} 
        \left( 1 - \left( \nh_{\A} \cdot \nh_{\B}\right)^{2} \right) .
\end{align}

If $c_{1}$ is to vanish identically, it must be the case that either each of 
the $c_{1,i}$ coefficients must equal zero independently, or $\Omega$ must take some special value such that the $\Omega$ polynomial vanishes.
The zeros of the $c_1$ polynomial in $\Omega$ are found with a straightforward application of the 
quadratic formula, and we find that the discriminant ($c_{1,1}^{2} - 4 c_{1,2} c_{1,0}$) is always negative 
for any choice of polarizations ($\nh_{\A / \B}$) or population fraction $N_\A/N$.  This implies that there exists 
no real value of $\Omega$ such that the coefficient $c_{1} = 0$.

Finally, we note that if $c_{1,2}$ vanishes, then both $\nh_{\A}$ and $\nh_{\B}$ must 
be parallel to $\pm \Bv$.  The coefficients $c_{1,1}$ and $c_{1,0}$ also vanish in this case, as they are both identically zero if $\nh_{\A} = \pm \nh_{\B}$.  We thus conclude that $c_{1}$ only vanishes when the initial polarization states are eigenstates of the one-body Hamiltonian. This is exactly the situation for all models studied in Refs.~\cite{roggero2021dynamical,roggero2021entanglement} where important deviations from mean-field behavior were shown to persist up to macroscopic system sizes.

We also calculate the skewness (denoted $\mathcal{M}_{3}$ below) and 
kurtosis (denoted $\mathcal{M}_{4}$ below) of the Hamiltonian with respect to this 
initial condition.  Both of these moments involve a large amount of algebra, and displaying 
the exact analytic expressions is prohibitively difficult. 
While we are able to calculate these moments 
using exact analytic expressions, their asymptotic behavior in the $N \rightarrow \infty$ 
limit is all that is necessary for obtaining some insight in the subsequent discussion.  
We find that
\begin{align} \label{eq:moms34}
    \lim_{N \rightarrow \infty} 
        \mathcal{M}_{3} &\equiv 
        \frac{\langle (H - \langle H \rangle)^{3} \rangle}{\Delta H^{3}} = 0 \\
    \lim_{N \rightarrow \infty} 
        \mathcal{M}_{4} &\equiv 
        \frac{\langle (H - \langle H \rangle)^{4} \rangle}{\Delta H^{4}} = 3 .
\end{align}
These limits are only violated when $c_{1}$ in $\Delta H^{2}$ is identically zero. 

The moment structure of the Hamiltonian when calculated with respect to our 
prototypical initial condition suggests that the probability density associated 
with measuring a given energy eigenstate with some nonzero overlap with our initial condition 
in the spectrum of the Hamiltonian is approximately a 
Gaussian distribution centered on $\langle H \rangle$ in the large $N$ limit. 
The width of the total energy spectrum of the Hamiltonian ($E_{\text{max}} - E_{\text{min}}$) 
scales proportionally with $N$, but the width of the initial condition in energy space 
scales like $\sqrt{N}$.  As $N$ becomes large, the Gaussian becomes (relatively) more localized 
in energy space. Thus our prototypical initial condition only accesses a fraction of the 
total spectrum of the Hamiltonian. As the energy spacing is $ \delta E \propto 1/N$, the approximate number 
of energy states which may (potentially) be accessed by the initial condition ($\mathcal{N}(\Psi)$) is 
\begin{equation}
\mathcal{N}(\Psi) \propto \frac{\sqrt{\Delta H^2}}{\delta E} \sim N^{3/2}
\end{equation}

The structure of the variance of the Hamiltonian, Eq.~\ref{eq:varH}, suggests a natural 
criterion for approximating the total number of neutrino flavor spins which must 
be included in a large scale many body simulation of a given system in order to isolate 
what features of the determined solution might persist in the large $N$ limit, and what 
features are finite size effects which will diminish with sufficiently large $N$.  We find that 
in cases such that $c_{1} N \gg c_{0}$ the evolution of simple one-body observables follows closely the one predicted by the mean field 
equations of motion on natural time scales predicted from the system parameters for several 
categories of initial conditions. Interestingly, for the cases investigated in this work, this convergence to the mean field evolution of one-body properties occurs despite the presence of a substantial fraction of the maximum entanglement (which is proportional to $\log(N)$ due to the size of the accessible Hilbert space) in the evolved many-body state (see Sec.~\ref{sec:entanglement}). This suggests that many-body quantum correlations generated by the time evolution are highly non-local in nature and might not be important to describe some aspects of the flavor dynamics in the system.

\section{Many-Body and Mean-Field Dynamics} 
\label{sec:dynamics}

In order to assess the impact of coherent neutrino flavor oscillations on the 
relevant astrophysical systems, we are 
interested in the expectation values of the one body operators $\langle \Jv_{\A/\B} \rangle$ 
which tell us about the average flavor content of the individual neutrino beams.  
The time behavior of these systems can be calculated in both
the full quantum evolution and in a mean-field (product state)
approximation. For the problem under our consideration the symmetries enable
us to calculate the full quantum evolution for thousands of spins
through the time evolution of the individual eigenstates obtained
as above.

Another method for following the evolution of the expectation values is through an application 
of the Ehrenfest theorem.  When applied we recover an equation of motion for the 
one body operators in terms of expectation values of two body operators.  Unfortunately, there 
exists no exact closure of this relationship, as the EoMs for the two body expectation values are 
functions of three body expectation values, and so
on~\cite{bogoliubov1946kinetic,Born:1946vqa,kirkwood1946statistical,yvon1935theorie,Volpe2013}.

In the mean-field approach, we approximate the two body expectation value as a product of one body 
expectation values with the goal of 
constructing a closed set of equations for the EoMs of the one 
body operator expectation values.  We therefore make the substitution
\begin{equation} \label{eq:twoBodyApprox}
    \langle \Jv_{\A} \times \Jv_{\B} \rangle \approx 
        \langle \Jv_{\A} \rangle \times \langle \Jv_{\B} \rangle .
\end{equation}
By defining polarization vectors as 
\begin{equation} \label{eq:pvecDef}
    \Pv_{\A/\B} = \frac{2}{N_{\A/\B}} \langle \Jv_{\A/\B} \rangle ,
\end{equation}
we recover the mean-field equations of motion:
\begin{subequations} \label{eq:meanFieldEOM}
    \begin{align}
        \frac{\dd \Pv_{\A}}{\dd t} &= \frac{\Omega}{2} \Bv \times \Pv_{\A}
             + \frac{N_\B}{N} \Pv_{\B} \times \Pv_{\A} \\
        \frac{\dd \Pv_{\B}}{\dd t} &= -\frac{\Omega}{2} \Bv \times \Pv_{\B}
             + \frac{N_\A}{N} \Pv_{\A} \times \Pv_{\B} .
    \end{align}
\end{subequations}

Naturally we are interested in whether or not these mean-field equations of motion are a reasonably 
accurate representation of the dynamics of the expectation values of the true spin-block-averaged one-body 
operators.  If we were able to show that is the case, Eq.~\ref{eq:meanFieldEOM} would represent a more efficient means 
of determining the expectation values of average one body operators despite its inherent nonlinearity.

The mean-field equations of motion above have been thoroughly studied, and we now present two primary categories 
of initial condition which we will utilize in our comparison between the mean-field and many-body results.
These solutions are compared with the full quantum many-body evolution
below.  These collective solutions, since they can be
obtained exactly in the full many-body system for hundreds
or thousands of spins, can also serve as tests and demonstrations
for quantum simulators including cold Rydberg atom arrays and
trapped ion systems.

\subsection{Mean-Field Collective Oscillations}
In order to directly compare the many-body and mean-field results, it is useful to have some qualitative insight 
into the behavior of the mean-field collective oscillations.  To this end, in this section we will present 
two primary categories of initial condition which have well understood dynamical evolution.  The first are 
bipolar initial conditions, characterized by the two polarization vectors initially in neutrino flavor 
states, thus begin anti-parallel and nearly aligned with $\Bv$.  The second are collective precession 
modes in which the two polarization vectors begin co-planar but not necessarily parallel.  These modes evolve 
by simply precessing about the $\Bv$ vector, and do not dynamically evolve along $\Bv$.

\subsubsection{Bipolar oscillations}
\label{ssec:bipolar_mf}

In the mass basis, $\Bv = -\ez$, and we will choose $\Pv_\B$ in the $\nu_{e}$ state, and $\Pv_\A$ in the 
$\nu_{x}$ state.  These initial conditions result in the famous ``bipolar" solutions to the mean-field EoMs.
It is well known that this category of solutions is isomorphic to a 
pendulum with a bob which itself has some internal angular momentum (known as a 
gyroscopic pendulum) \cite{Hannestad:2006nj}. The initial polarization vectors corresponding to neutrino flavor states take the form
\begin{align} \label{eq:mfBipIC}
    \Pv_{\B / \A}(t=0) \rightarrow &\Pv_{e/x}(t=0) =  \nonumber \\
        &\pm \left( \svac \ex + \cvac \ez \right) 
\end{align}
where $\tvac$ represents the mixing angle between neutrino flavor and mass eigenstates, and $+(-)$ is chosen for the initially 
$\nu_{e}(\nu_{x})$ flavor beam. 

This system of equations experiences significant excursions away from the initial condition only for certain 
values of the population fractions ($N_{e / x}/N$) and the vacuum oscillation frequency $\Omega$. To 
see this, we assume that $\svac \ll 1$ due to suppression from the large local matter density, and we then linearize 
the equations of motion in terms of the small (complex) variable
\begin{equation}
    \Pv_{e/x} \cdot \left( \ex - \I \ey \right) = \epsilon_{e/x} .
\end{equation}
If the system is to experience significant flavor conversion, it must be true that 
$|\epsilon_{e/x}|$ grows significantly.  As long as $|\epsilon_{e/x}|$ is small, we can 
find the eigenvalues of the linearized EoMs, and when these eigenvalues become complex 
for some choice of $\Omega$ and $N_e/N$ the system will experience exponential 
growth in $|\epsilon_{e/x}|$.  It can be shown that the range of parameters for which significant 
flavor oscillations can occur is given by the inequality
\begin{equation} \label{eq:bipInstability}
    \left( \sqrt{\eta} - \sqrt{1-\eta} \right)^{2} < \Omega < 
        \left( \sqrt{\eta} + \sqrt{1-\eta} \right)^{2}
\end{equation}
where we have defined $\eta = N_e / N$ as the fraction of the total spins initially in the ($e$)lectron neutrino 
flavor state. (See \cite{Banerjee:2011fj} for a more comprehensive discussion of the stability of dense neutrino gases.) 

We can compute the variance of the many-body Hamiltonian with respect to the bipolar-type initial polarizations and we find that
\begin{equation}
    \Delta H^{2} = \left( \frac{\Omega \svac}{4} \right)^{2} N + \eta (1-\eta) .
\end{equation}
Our hypothesis is that when $c_{1} N \gg c_{0}$, we should expect to see behavior emerge in the many-body 
system which will persist in the large $N$ limit, and which may correspond to solutions found using the 
mean-field equations of motion.  We observe that there is a tension which occurs between these 
two formalisms.  The mean-field equations of motion predict that significant excursions from the initial 
condition are only possible for some range of $\Omega$ given in the inequalitities of Eq.~\ref{eq:bipInstability}, and for $\Omega$ 
outside this interval flavor excursions can occur on finite timescales for finite values of $N$ in the many-body formalism \cite{roggero2021entanglement}. This 
inequality pins $\Omega$ to be of $\mathcal{O}(\eta)$ 
(Except in the case that $\eta = 0.5$ in which case $\Omega \rightarrow 0$ is also predicted to be unstable).
Given that $\svac \ll 1$, we generically expect that $c_{0}/c_{1} \gg 1$ which subsequently implies $N$ must be such that $N \gg 1/\svac^{2}$ to see the emergence of mean-field like behavior in a full many-body calculation.

\begin{figure}
 \includegraphics[width=0.48\textwidth]{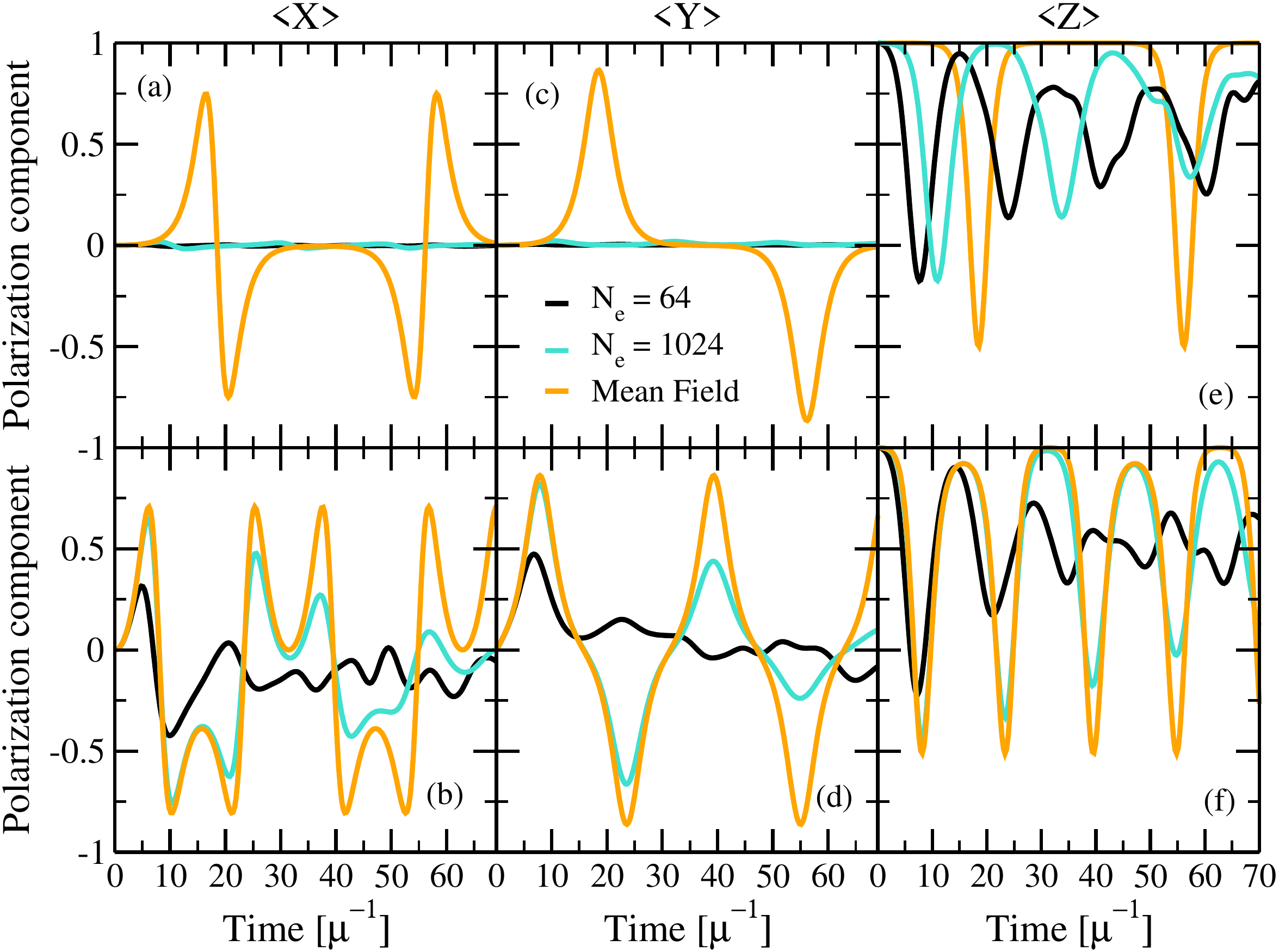}
 \caption{Evolution of the 3 components of the polarization vector as a function of time in both the mean field approximation (orange line) and many-body simulations with $N_e=N_x=64$ (black line), $N_e=N_x=1024$ (turquoise line) and $\Omega = 0.5$ (all lines and panels). Panels (a-b) show the $x$ component, panels (c-d) the $y$ component and panels (e-f) the $z$ component of the polarization vector. The top row (panels a,c,e) are for a small mixing angle $\theta=0.001$ ($c_0/c_1 \approx 4.0 \times 10^{6}$)  while the bottom row (panels b,d,f) use a large mixing angle $\theta=0.1$ ($c_0/c_1 \approx 4.054 \times 10^{2}$). Note that these two results correspond to cases 3 and 7 in the next subsection.}
\label{fig:xyz_bipolar}
\end{figure}

We present an example of the dynamic behavior of bipolar mode oscillations in Fig.~\ref{fig:xyz_bipolar} for 
two choices of the mixing angle $\tvac$, equal numbers of initially $\nu_e$ and $\nu_x$ flavor neutrinos, and $\Omega = 0.5$. (This choice of parameters corresponds to cases 3 and 7 in table \ref{tab:bipolar} of the next subsection.) We find that in the case $\tvac \ll 0.1$, the evolution of the many-body solution departs quickly and significantly from the behavior of the mean-field.  In this case, increasing $N$ lengthens the time at which a significant difference accumulates in the polarization vectors calculated in the two formalisms.  When $\tvac$ is $\mathcal{O}(.1)$ we find that for $N=128$ there is already significant disagreement between the two formalisms before the minimum of the first major oscillation.  However by increasing the number of spins to $N=2048$ we achieve $c_0/c_1 N \approx 0.2 \ll 1$, and we begin to see the mean-field behavior emerge in the one-body expectation values of the many-body solution.

\subsubsection{Collective precession}
\label{ssec:precession_mf}

Another evolution mode which arises from the mean field EoMs are those such that there is no dynamic 
evolution along $\ez$, but the polarization vectors merely precess 
in the $\ex$ - $\ey$ plane. Such solutions require that the polarization 
vectors all be coplanar initially and precess with the same oscillation frequency 
denoted $\Omega^{c}$ \cite{Raffelt:2007cb,Raffelt:2007xt,Xiong:2021dex,Duan:2021woc}.

We can find such solutions for our mean-field two-block example system by 
first taking the ansatz
\begin{equation}
    \Pv_{\A/\B} = \begin{pmatrix}
        \sin(\theta_{\A/\B}) \cos(\Omega^{c} t) \\
        \sin(\theta_{\A/\B}) \sin(\Omega^{c} t) \\
        \cos(\theta_{\A/\B}) \end{pmatrix}  .
\end{equation}
In order that these ansatz polarization vectors satisfy the 
equations of motion, the two polar angles $\theta_{\A/\B}$ must satisfy the (nonlinear) system of equations
\begin{align} \label{eq:precessReqs}
    L &= \frac{N_\A}{N} \cos(\theta_{\A}) + \frac{N_\B}{N} \cos(\theta_{\B}) \\
    \Omega \sin(\theta_{\A}) \sin(\theta_{\B}) &=
        \sin(\theta_{\A} - \theta_{\B}) \times \nonumber \\
        &\quad    \left( \frac{N_\A}{N} \sin(\theta_\A)+\frac{N_\B}{N} \sin(\theta_\B)\right)
\end{align}
Here, the quantity $L$ represents the fractional population difference between initially $\nu_{1}$
neutrinos and initially $\nu_{2}$ neutrinos summed over both beams which 
is a quantity conserved by both the many-body and mean-field equations of motion.  In what follows, 
we will choose it somewhat arbitrarily in order to investigate a range of values for $\Omega^c$.

Given that we know $L$, $\Omega$, and the population fraction of spins in each block, we can find the 
polarization angles ($\theta_{\A / \B}$) necessary such that our ansatz initial condition will precess with 
frequency $\Omega^{c}$.  By taking a time derivative of $\Pva \cdot \hat{e}_{1}$, we can find an explicit 
expression for $\Omega^{c}$ in terms of the initial condition and physical parameters.  
We can do the same with $\Pvb$, and find two equations which are equal by assumption.  
These are:
\begin{align} \label{eq:precessionFreqs}
    \Omega^{c} &= -\frac{\Omega}{2}
        + \frac{N_\B}{N} \frac{\sin(\theta_\A - \theta_\B)}{\sin(\theta_{\A}) } \\
    \Omega^{c}  &= \frac{\Omega}{2}
        -  \frac{N_\A}{N} \frac{\sin(\theta_\A - \theta_\B)}{\sin(\theta_{\B})}
\end{align}

As in the bipolar case, we can also compute the variance of the many-body Hamiltonian for initial polarizations which 
result in collective precession modes.  We find that the values of $c_{1}$ and $c_{0}$ take the form
\begin{align}
    c_{0} &= \frac{N_{\A} N_{\B}}{4N^{2}} \left( 1 - \cos(\theta_\A - \theta_\B) \right)^{2} \\
    c_{1} &= \left(\frac{\Omega^{c}}{2}\right)^{2} \left( 
         \frac{N_\A}{N} \sin^{2}(\theta_{\A}) +  \frac{N_\B}{N} \sin^{2}(\theta_{\B})
    \right) .
\end{align}
While $c_{0}$ follows straightforwardly from Eq.~\ref{eq:c0}, we find that $c_{1}$ is proportional to the square 
of the collective precession frequency.  This then provides an ideal playground for testing for our hypothesis, as 
we can simply find precession solutions which precess slowly or rapidly in order to maximize or minimize $c_{0}/c_{1}N$.

\begin{figure}
 \includegraphics[width=0.48\textwidth]{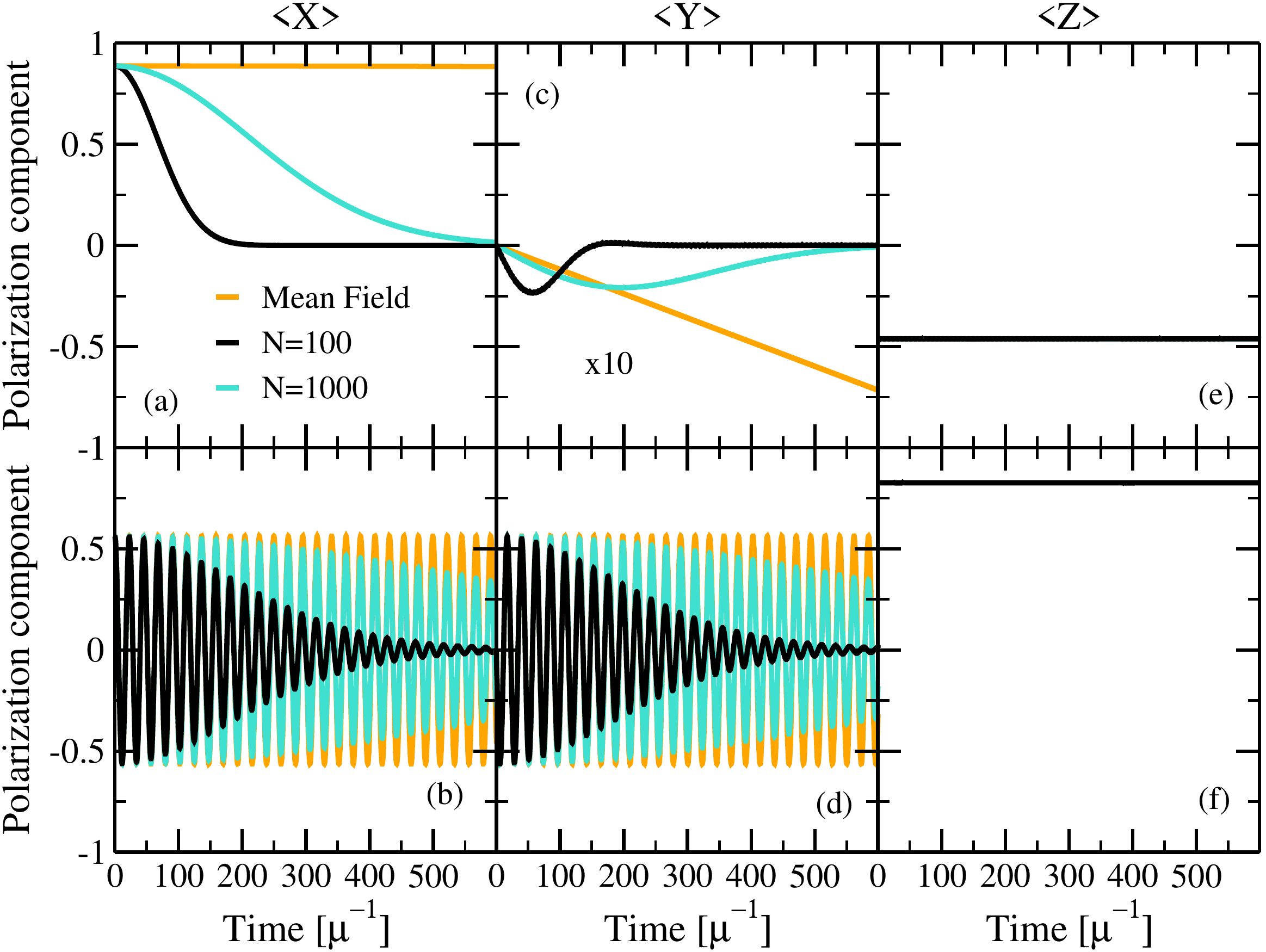}
  \caption{ Evolution of the 3 components of the polarization vectors describing beam $\A$ for two precession mode solutions.  The top panels represent a precession mode with $\Omega^{c} \approx 1.35 \times 10^{-4}$ ($c_0/c_1 \approx 1.25 \times 10^{6}$), and the bottom panels have $\Omega^{c} \approx 2.77 \times 10^{-1}$ ($c_0/c_1 \approx 9.018 \times 10^{-2}$). The black and cyan curves employ $N = 100$ and $N = 1000$ total spins respectively. For the larger precession frequency, $c_0/c_1 N \ll 1$ may be obtained with many fewer spins than for the lower precession frequency, and the qualitative correctness of the mean-field prediction is maintained over longer time periods by increasing $N$ once $c_0/c_1 N \ll 1$ is reached.  The plotted curves utilize the parameters in cases $7$ (top panels) and $1$ (bottom panels) provided in table \ref{tab:precession}.}
\label{fig:xyz_precession}
\end{figure}

We present two examples of collective precession mode solutions in both the many-body and mean-field approximation in 
Fig.~\ref{fig:xyz_precession} with two precession oscillation frequencies which differ by three orders of magnitude. In the top panel, we show a 
precession mode with a precession frequency of $\Omega^{c} \approx 1.35 \times 10^{-4}$, while the bottom panel precesses with 
frequency $\Omega^{c} \approx 2.77 \times 10^{-1}$.  As $c_0/c_1$ is approximately $7$ orders of magnitude larger for the top panels, 
we expect to need a proportionally larger number of spins to see the emergence of mean-field like behavior for that choice of parameters.  
However, we do note that the many-body solution still respects the precession requirement that the polarization vectors do not 
dynamically evolve along $\Bv$ even in the case that there is no transient agreement between the mean-field and many-body formalisms (top panels of Fig.~\ref{fig:xyz_precession}).  We leave the investigation of this intriguing behavior for future work.

\subsection{One-body Observables and Mean-Field Emergence in Many-Body Solutions}
\label{ssec:mf_from_manybody}

In order to demonstrate the large $N$ behavior of the many-body system and compare it with the mean-field solutions, we investigated 
seven cases in each of three categories of solution to the many-body and mean-field neutrino oscillations problem. We then inspect 
the deviation between the large $N$ many-body calculations, and the mean-field approximation. We concentrate on one-body observables like the
flavor content versus beam (more generally energy and angle), which
are the observables that can be detected in terrestrial neutrino 
observatories.

The first class of solution are bipolar modes, the mean-field configurations and solutions of which are described in the previous subsection.  For the different choices of physical parameter, we solved the corresponding many-body problem with 
$N = [700,1000,1300,1600,3600] $.  The second are collective precession solution modes, also as 
described previously.  Finally we randomly chose the values of $\Omega \in (-3.0,3.0)$ and $N_{\A}/N \in (0,1.0)$ as well as the 
polarizations of the two spin blocks, $\nh_{\A / \B}$.  For all of the precession modes, and random parameter calculation sets we chose 
$N = [ 100, 200, 300, 400, 800 ]$.  We used significantly more spins in the bipolar cases in light of our insight that the bipolar modes 
represent very sparse distributions in energy space, and a sufficiently large $N$ is necessary for the energy distribution to be approximately 
Gaussian.

For every choice of parameters and total number of spins ($N$), we solved the systems to a time
\begin{equation}
\label{eq:tf}
    t_{f} = 3 \sqrt{\frac{N}{\Delta H^{2}}} = \frac{3}{\sqrt{c_1}}\left( \frac{1}{\sqrt{1 + \frac{c_0}{c_{1}N}}} \right).
\end{equation}
This time has an $N$-independent term and a correction term resulting from the finite size of the many-body system. This correction goes to zero in the large $N$ limit, and thus this time represents a natural scale over which the system's size-invariant behavior in the many-body case may be expected to be observed.  As this is a persistent timescale as $N$ becomes large, it is natural to employ in comparisons with mean-field evolution.  We do indeed observe that this timescale is directly proportional to the inverse of the collective precession frequency of the mean-field precession modes.

Furthermore, in the large system size limit it is expected that at some finite time the mean-field and many-body predictions for the evolution of the one body operators will diverge.  Because the above time
approaches a constant as $N$ becomes large, it does not represent this divergence time which should depend on $N$. Because of its invariance for sufficiently large systems and our observation that it is directly related to the evolution timescale of at least one mean-field evolution mode, we employ this timescale for evolving our systems in order to self-consistently compare results between different oscillation modes and parameter choices which evolve on significantly differing timescales.

As a simple measure of agreement between the mean-field and many-body single ``beam" observables, we define the polarization vectors in the many-body system 
according to Eq.~\ref{eq:pvecDef} and calculate the magnitude of the vector difference between the many-body and mean-field $\Pv$ vectors for both 
block $\A$ and block $\B$.  We then take the largest magnitude value which occurs in both spin blocks of this vector difference over our solution interval. 

\begin{figure}
    \begin{center}
        \includegraphics[scale=0.4]{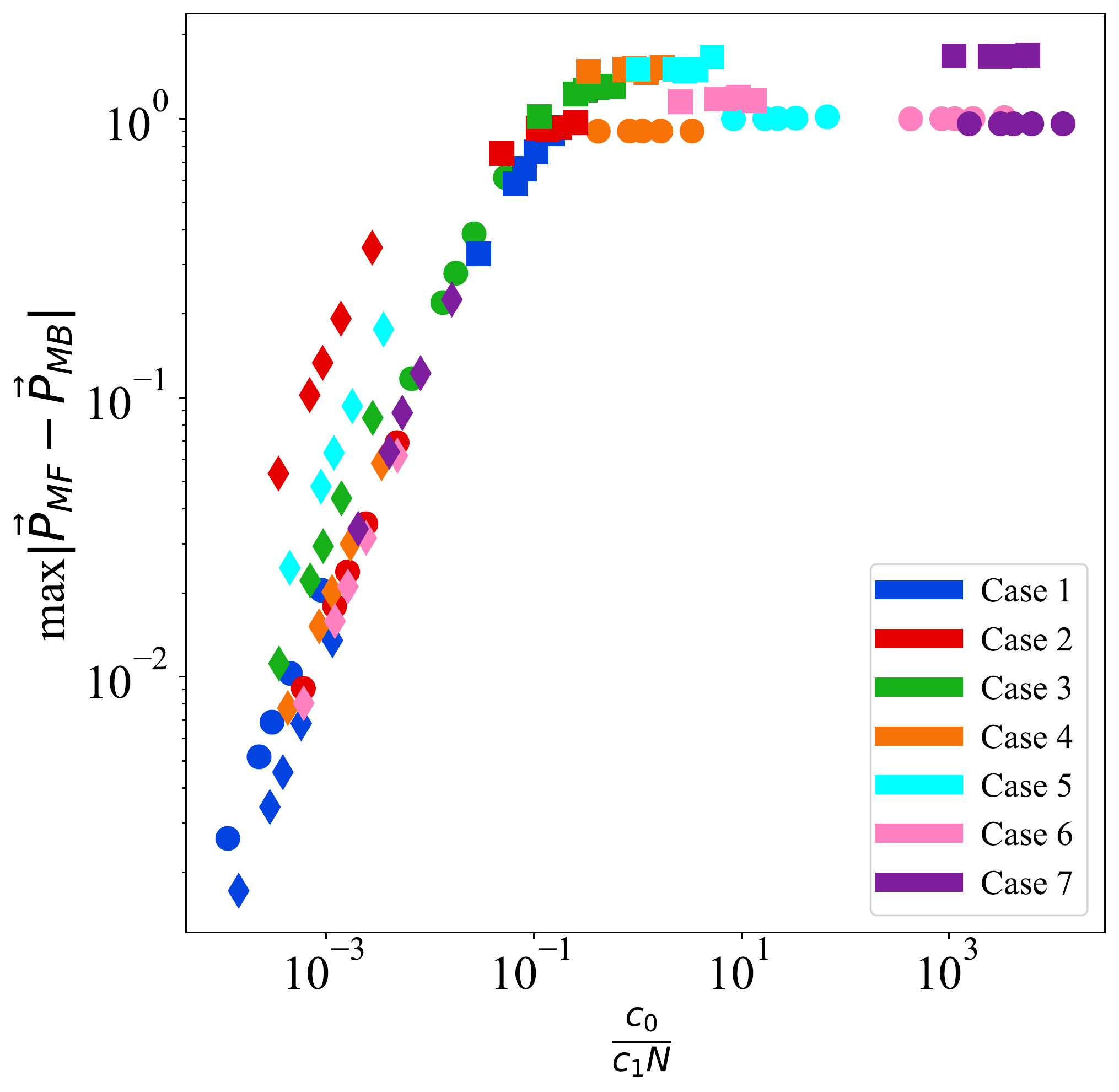} 
    \end{center}
    \caption{Each marker indicates the largest difference between the many-body (MB) and mean-field (MF) polarization vectors for a choice of solution mode (marker shapes), and each color represents a choice of parameter set tabulated in one of the tables \ref{tab:bipolar},\ref{tab:precession}, and \ref{tab:random}.  Squares represent bipolar mode solutions, circles represent collective precession modes, and diamonds represent 
    solutions for randomly chosen parameters.  The multiplicity of markers is due to increasing values of $N$, with increasing $N$ from right to left in a given marker shape and color.}
    \label{fig:mf_v_mb}
\end{figure}

In Fig.~\ref{fig:mf_v_mb} we show the behavior of the largest deviation in polarization between the many-body and corresponding mean-field behavior 
for each $N$ and parameter set.  In this figure, color denotes the choice of physical parameters, tabulated by case number (\# column) in tables \ref{tab:bipolar} through 
\ref{tab:random}.  Square markers represent bipolar mode solutions, circle markers represent precession mode solutions, and diamonds represent randomly chosen 
parameters.  We note that there is a categorical difference in solution behavior in the regime in which $c_0/c_1 N \gg 1$, and the case such that 
$c_0/c_1 N \ll 1$.  For all cases in which $c_0/c_1 N \gg 1$, the deviation between many-body and mean-field solutions is approximately maximal on our solution interval, 
but when $c_0 / c_1  N$ approaches $0.1$ there is a knee in the deviation of the solutions past which increasing $N$ results in improving agreement between the 
many-body and mean-field approaches.
Thus we observe that the ratio condition $c_0/c_1 \approx N$ can be used as a heuristic for determining the number of spins which must be included to observe mean-field like behavior emerge in full many-body calculations, however these curves are not identical so further refinements on a per-calculation-basis are still 
required to demonstrate full convergence of one-body observables.

\begin{table}[h!] 
    \begin{tabular}{ | p{0.4cm}||p{0.9cm}|p{0.8cm}|p{1.3cm}|p{0.4cm}|p{0.8cm}|p{0.4cm}|p{1.6cm}| }
    \hline
    \# & $N_e/N$ & $\Omega$ & $\theta_{\A}$ & $\phi_{\A}$ & $\theta_{\B}$ & $\phi_{\B}$ & $ c_{0} / c_{1} $  \\
    \hline
    \hline
    \textbf{1} & $0.5$ & $0.5$ & $\pi - 0.4$ & $\pi$ & $0.4$ & $0.0$ & $1.06 \times 10^2$ \\
    \hline
    \textbf{2} & $0.55$ & $1.5$ & $\pi - 0.1$ & $\pi$ & $0.1$ & $0.0$ & $1.77 \times 10^2$ \\
    \hline
    \textbf{3} & $0.5$ & $0.5$& $\pi - 0.2$ & $\pi$ & $0.2$ & $0.0$ & $4.05 \times 10^2$ \\
    \hline
    \textbf{4} & $0.25$ & $0.5$ & $\pi - 0.1$ & $\pi$ & $0.1$ & $0.0$ & $1.20 \times 10^3$ \\
    \hline
    \textbf{5} & $0.55$ & $0.33$ & $\pi - 0.1$ & $\pi$ & $0.1$ & $0.0$ & $3.65 \times 10^3$ \\
    \hline
    \textbf{6} & $0.75$ & $0.18$ & $\pi - 0.1$ & $\pi$ & $0.1$ & $0.0$ & $9.29 \times 10^3$ \\
    \hline
    \textbf{7} & $0.5$ & $0.5$ & $\pi - 0.002$ & $\pi$ & $0.002$ & $0.0$ & $4.00 \times 10^6$ \\
    \hline
    \end{tabular}
    \caption{Parameters utilized in bipolar mode solutions to both the many-body and mean-field EoMs. $N_e/N$ and $\Omega$ were chosen arbitrarily but satisfy the inequality in
        Eq.~\ref{eq:bipInstability}.  Finally, the table is ordered by ascending values of $c_0/c_1$.}
    \label{tab:bipolar}
\end{table}

\begin{table}[h!]
    \begin{tabular}{ | p{0.3cm}||p{0.9cm}|p{0.6cm}|p{1.4cm}|p{0.5cm}|p{1.4cm}|p{0.5cm}|p{1.6cm}| }
    \hline
     \# & $N_\A/N$ & $\Omega$ & $\theta_{\A}$ & $\phi_{\A}$ & $\theta_{\B}$ & $\phi_{\B}$ & $ c_0 / c_1 $  \\
    \hline
    \hline
    \textbf{1} & $0.51$ & $1.2$ & $0.5978067$ & $0.0$ & $0.2175694$ & $0.0$ & $0.0902 $ \\
    \hline
    \textbf{2} & $0.45$ & $1.5$ & $1.050692$ & $0.0$ & $0.2942370$ & $0.0$ & $0.482$ \\
    \hline
    \textbf{3} & $0.51$ & $0.2$ & $1.443493$ & $0.0$ & $1.248403$ & $0.0$ & $5.32$ \\
    \hline
    \textbf{4} & $0.48$ & $0.9$ & $2.012938$ & $0.0$ & $1.079368$ & $0.0$ & $3.34 \times 10^{2}$ \\
    \hline
    \textbf{5} & $0.27$ & $1.37$ & $1.568292$ & $0.0$ & $0.3723205$ & $0.0$ & $6.70 \times 10^3$ \\
    \hline
    \textbf{6} & $0.33$ & $1.2$ & $1.618388$ & $0.0$ & $0.5131689$ & $0.0$ & $3.40 \times 10^5$ \\
    \hline
    \textbf{7} & $0.52$ & $0.75$ & $2.051478$ & $0.0$ & $1.286571$ & $0.0$ & $1.25 \times 10^6$ \\
    \hline
    \end{tabular}
    \caption{Parameters utilized in collective precession mode solutions to both the many-body and mean-field EoMs. 
    Parameters were chosen to span a wide range of $\Omega^{c}$, but were otherwise taken arbitrarily.  They are presented with seven significant figures. As in the bipolar case, the table is ordered by ascending values of $c_0/c_1$.}
    \label{tab:precession}
\end{table}

\begin{table}[h!] 
    \begin{tabular}{ | p{0.4cm}||p{0.9cm}|p{1.0cm}|p{0.9cm}|p{0.9cm}|p{0.9cm}|p{0.9cm}|p{1.1cm}| }
    \hline
    \# & $N_\A/N$ & $\Omega$ & $\theta_{\A}$ & $\phi_{\A}$ & $\theta_{\B}$ & $\phi_{\B}$ & $ c_0 / c_1 $  \\
    \hline
    \hline
    \textbf{1} & $0.66$ & $-1.396$ & $2.920$ & $4.854$ & $2.386$ & $2.027$ & $0.115$ \\
    \hline
    \textbf{2} & $0.81$ & $0.3134$ & $1.972$ & $4.179$ & $2.771$ & $5.550$ & $0.277$ \\
    \hline
    \textbf{3} & $0.18$ & $-1.859$ & $0.4564$ & $1.451$ & $1.278$ & $4.236$ & $0.280$ \\
    \hline
    \textbf{4} & $0.49$ & $1.464$ & $0.8108$ & $3.545$ & $0.3045$ & $0.1005$ & $0.342 $ \\
    \hline
    \textbf{5} & $0.83$ & $2.371$ & $2.339$ & $2.258$ & $1.133$ & $3.828$ & $0.356$ \\
    \hline
    \textbf{6} & $0.45$ & $1.5$ & $1.881$ & $4.263$ & $2.175$ & $2.174$ & $0.485$ \\
    \hline
    \textbf{7} & $0.29$ & $2.032$ & $1.996$ & $0.6419$ & $0.5526$ & $3.708$ & $1.62$ \\
    \hline
    \end{tabular}
    \caption{Parameters chosen at random (except case 6 for which only the polarization angles were chosen at random.  The population fractions and $\Omega$ were chosen to match cases 
    1 and 2 in the bipolar and precession modes respectively).  We chose $\Omega \in (-3.0,3.0)$, $N_\A/N \in (0,1.0)$, 
    each polar angle $\theta_{\A/\B} \in (0,\pi)$, and each azimuthal angle 
    $\phi_{\A/\B} \in (0,2\pi)$, and values are specified with four significant figures.  As in the previous two solution categories, this table is ordered by 
    ascending values of $c_0/c_1$.}
    \label{tab:random}
\end{table}

We also note the qualitative differences in the convergence behaviors of the three categories of flavor oscillations.  Bipolar mode oscillations are characterized by large values of $c_{0}/c_{1}$ due both to the smallness of $\sin^{2}(2 \theta)$ in the denominator of this ratio and the limit placed on $\Omega$ by the inequality of Eq.~\ref{eq:bipInstability}.  Precession mode solutions display a wide range of values for $c_{0}/c_{1}$ determined by the collective precession frquency $\Omega^{c}$.  When the initial polarizations, population fractions and vacuum oscillation frequency are chosen at random, the ratio $c_{0}/c_{1}$ is close to $\mathcal{O}(1)$ implying that in the majority of arbitrary cases will only require a moderate number of total spins $N$ to observe behavior in agreement with an equivalent mean-field calculation.

\subsection{Special cases}
As mentioned above, the model in Eq.~\eqref{eq:twoBeamH} can be solved numerically exactly using the Bethe ansatz but for particular choices of the parameters the full time-evolution can be computed analytically. For models without the vacuum contribution, ie. setting $\Omega=0$, and for $N_A=N_B$ with $\theta_\A=0$ and $\theta_\B=\pi$, Friedland and Lunardini have shown how to express the evolved state analytically in terms of appropriate Clebsh-Gordan coefficients~\cite{Friedland2003}. In this limit, the mean-field equation of motion Eq.~\eqref{eq:meanFieldEOM} predict no flavor evolution while the full many-body treatment shows significant oscillations. The time-scale for these oscillations grows however very quickly with system size as $t=\mathcal{O}(\sqrt{N_\A})$ and quickly diverges for large systems, indicating that the mean-field prediction is qualitatively correct. These results were later extended in Ref.~\cite{Friedland2006} to the asymmetric case $N_\A\neq N_\B$ and it was shown that the amplitude of oscillations in these models decays as a polynomial in $|N_\A-N_\B|$, once again showing the qualitative correctness of the mean-field approximation. In recent work by one of us~\cite{roggero2021entanglement} it was shown, using a many-body simulation employing Matrix Product States(MPS), that a system with $N_\A=N_\B$ starting in a product state can develop an entanglement entropy scaling as $S=\mathcal{O}(\log(N_\A))$ showing that there exist observables which will fail to be predicted correctly in a mean-field calculation (which by construction have $S=0$ at all times.) We will investigate the behavior of entanglement measures in the next section.

\section{Entanglement as order parameter for instability}
\label{sec:entanglement}

The use of entanglement measures to characterize different phases of matter and to classify many-body states in terms of their correlation structure and topological properties has a long history in condensed matter physics (see e.g~\cite{Osterloh_2002,Vidal_qcp_2003,Plenio2005}) and more recently has been applied to systems in nuclear and high-energy physics producing interesting insights (see e.g.~\cite{Klebanov:2007ws,Beane:2018oxh,Beane:2019loz}). In the context of collective neutrino oscillations the role of quantum correlations is not fully understood yet, on one hand entanglement has been associated with a speed-up of flavor conversion~\cite{Bell:2003mg,sawyer2004classical} and on the other hand has been argued to not play any role in neutrino systems that are prepared in a mean-field state~\cite{Friedland2003b,Friedland2003}. Recent work adopting Tensor Network methods has shown how these, seemingly conflicting, results could be reconciled adopting the point of view that many-body coherent speed-up of flavor dynamics are generated when the neutrino systems under study undergoes a Dynamical Phase Transition~\cite{roggero2021entanglement,roggero2021dynamical}. In particular the scaling of entanglement with the size of a neutrino systems has been shown to be a strong indicator for the presence of bipolar modes~\cite{roggero2021entanglement,roggero2021dynamical} suggesting it could possibly be employed in conjunction with linear stability analysis to detect instabilities in a neutrino system. Recent work employing exact diagonalization techniques in small neutrino systems has also shown how entanglement can signal the presence of spectral splits in the neutrino spectrum~\cite{patwardhan2021spectral} (see also~\cite{Cervia2019,Rrapaj2020} for studies of entanglement in small neutrino systems and~\cite{xiong2021manybody} for an extension of the work in~\cite{roggero2021entanglement} to larger system sizes).

In this section we generalize the results presented in Refs.~\cite{roggero2021entanglement,roggero2021dynamical} to the more general bipolar case described by the Hamiltonian in Eq.~\eqref{eq:twoBeamH}. The crucial difference is that, for values of the mixing-angle $\theta_A$ and $\theta_B$ different from integer multiples of $\pi$, the mean-field approximation also predicts flavor evolution. Throughout this section when considering bipolar modes we will denote $\theta_B=\theta_A+\pi$ and simply denote $\theta_A/2$ as $\theta$.

\begin{figure}
 \centering
 \includegraphics[width=0.48\textwidth]{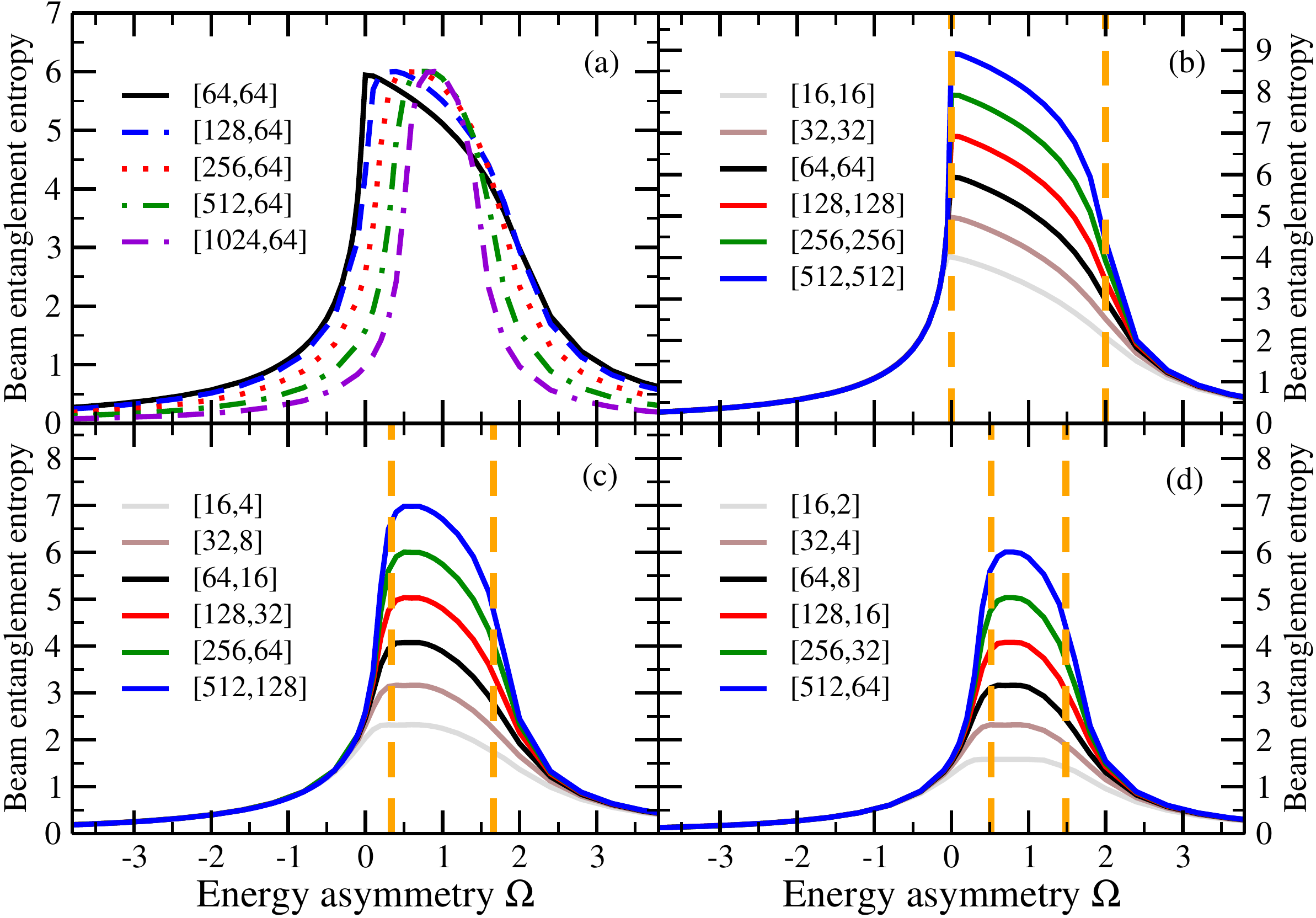}
 \caption{Entanglement entropy when the system is divided into the two beams. Panels (b-d) show the beam entanglement entropy as a function of the energy asymmetry $\Omega$ for three different population asymmetries $\eta=N_e/N$: panel (b) uses $\eta=1/2$, panel (c) uses $\eta=4/5$ and panel (d) is for $\eta=8/9$. Panel (a) instead shows the evolution of the beam entanglement entropy for fixed $N_x=64$ for different values of $N_e$ (corresponding to increasing $\eta$.) The vertical orange dashed lines correspond to the threshold values of the bipolar instability obtained from the analysis of the mean-field approximation in Eq.~\eqref{eq:bipInstability}.}
\label{fig:ent}
\end{figure}

We used the strategy described after Eq.~\eqref{eq:oneSpin} (exploiting the high degree of symmetry of the system and the sparsity of the Hamiltonian) to simulate large systems beyond the reach of Matrix Product State (MPS) simulations (limited to $N=\mathcal{O}(100)$ on a workstation) and report the results for the value of the maximum entanglement entropy in the minority beam in Fig.~\ref{fig:ent}. We have checked, using MPS simulations on systems with $N=64$ and various population asymmetries, that for these models this is indicative of the maximum bipartite entropy in the system.
Panels (b-d) of Fig.~\ref{fig:ent} show the entanglement entropy, defined as $S(N_{min}) =-$Tr$(\rho \log_2(\rho))$, where $\rho$ is the reduced density matrix for the beam with the fewest flavor spins, as a function of the energy asymmetry $\Omega$ for increasing values of the occupation $N_{min}$ in the minority beam while keeping the same ratio $N_{min}/N$ constant ($N_{min}/N=1/2,1/5,1/9$ for panels $(b)$,$(c)$ and $(d)$ respectively). In all of these results the value of the mixing angle was chosen to be $\theta=0.001$; for larger values of $\theta$ the behavior is similar but the transition is less sharp, and a similar behavior is found with the purity discussed below. The orange vertical dashed lines correspond to the boundaries of the unstable bipolar region from Eq.~\eqref{eq:bipInstability} which was obtained from the stability analysis of the mean field solution. The results for $N_{min}/N=1/2$ in panel (b) of Fig.~\ref{fig:ent} exactly match the earlier results from Refs.~\cite{roggero2021entanglement,roggero2021dynamical} and obtained using the MPS ansatz: inside the unstable region the entropy scales as $\log(N_{min})$ while outside we find a constant entropy for all the considered system sizes. The scaling of the entropy as $\log(N_{min})$ can be understood as a consequence of of the $N^{3/2}$ size of the Hilbert space accessible by the prototypical initial condition Eq.~\ref{eq:genIC} as discussed in sec.\ref{sec:moments}. The same behavior of the entropy is also found for different systems with a larger asymmetry between the occupation in the two beams (panels (c-d) of Fig.~\ref{fig:ent}) even though the left transition point becomes less sharp. The independence of the entanglement entropy on the size of the majority beam is shown in panel (a) of Fig.~\ref{fig:ent} where we show results for fixed $N_{min}=64$ and increasing occupation asymmetries ($N_{min}/N=1/2,1/3,1/5,1/9,1/17$). In all cases we find the same entanglement signature of the instability as in the other panels. For larger mixing angles the behavior of the entanglement entropy is similar with the only main difference that, at least for small system sizes, the curves as a function of $\Omega$ are less smooth and show small amplitude oscillations around the value found in the small mixing angle limit.

Another measure that shows entanglement production in the unstable region is the single neutrino purity, defined as follows~\footnote{Note that the definition used here differs from the standard one where the purity is simply $\text{Tr}[\rho_i^2]$.}
\begin{equation}
\begin{split}
\label{eq:def_purity}
\mathcal{P}_i &= \langle \sigma^x_i\rangle^2+\langle \sigma^y_i\rangle^2+\langle \sigma^z_i\rangle^2\\
&=2\text{Tr}[\rho_i^2]-1\;,
\end{split}
\end{equation}
with $\sigma^k_i$ the $k$-th Pauli matrix acting on the amplitude of the $i$-th neutrino in the system and $\rho_i$ its density matrix. In a mean field calculation we always have $\mathcal{P}_i=1$ at all times since this quantity measures the norm of the polarization vector. Using the results for the entropy discussed above, we can place an upperbound on the possible purity in the presence of non-zero entanglement. As we show in detail in Appendix~\ref{app:purity bound}, if we denote by $S_\A$ the entropy of beam $\A$ with $N_\A$ amplitudes and with $\mathcal{P}_\A$ the purity of a neutrino in that beam, we have (for $S_\A<N_\A/(4\ln(2))$)
\begin{equation}
\label{eq:purity_bound}
\mathcal{P}_\A < 1-e\left(\frac{S_\A}{N_\A}\right)^{\ln(2)}2^{-\sqrt{2\ln\left(\frac{N_\A}{S_\A}\right)}}\;.
\end{equation}
This shows that, even in systems with logarithmically scaling entropy $S_\A=\mathcal{O}(\log(N_\A))$, the individual neutrino flavor spins could be be found in an essentially pure state in the large $N_\A$ limit. A stronger argument would however require also a lower bound on the purity which in general we suspect cannot be obtained purely based on entropic arguments due to the subadditivity of the entanglement entropy. 

\begin{figure}
 \centering
\includegraphics[width=0.48\textwidth]{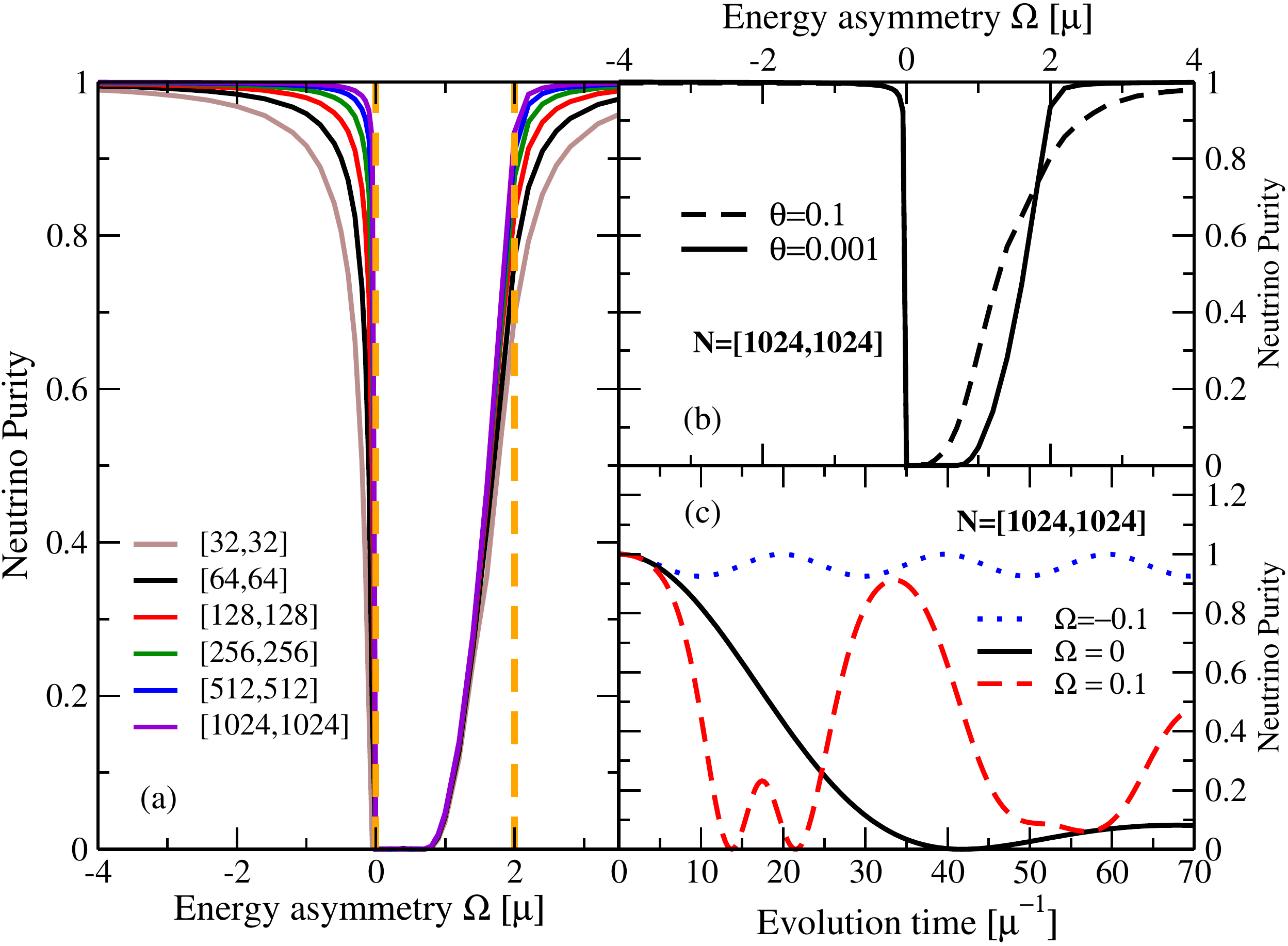}
 \caption{Panel (a) shows the minimum purity in simulations up to time $t\mu=70$ for neutrino system with $N_x=N_e$, $\theta=0.001$ and increasing system size. Panel (b) shows the minimum purity for the largest system and two different mixing angles $\theta=0.1$ (dashed line) and $\theta=0.001$ (solid line). Panel (c) shows the time evolution of the purity for a neutrino system with $N_x=N_e=1024$, $\theta=0.001$ and different values of the energy asymmetry.}
\label{fig:purity}
\end{figure}

We show several results for the purity in Fig.~\ref{fig:purity} for systems with equal numbers of neutrinos in the two beams, and, due to symmetry coming from $N_x=N_e$, we will consider only the purity evaluated for the first neutrino. In panel (a) we show the dependence of the minimum neutrino purity 
on the energy asymmetry $\Omega$ for systems of neutrinos of increasing size and at a small value ($\theta=0.001$) for the mixing angle. The unstable region is denoted by the vertical dashed orange lines and we can see that the purity deviates significantly from one in this region. As the system size increases the boundaries become sharper but we do not observe a significant increase in the value of the purity inside this region (possibly due to the system sizes being too small). Outside the unstable region we found a rapid convergence towards the mean field result with $1-P\approx\mathcal{O}(1/N)$ which is consistent with the entropy being constant outside the unstable region (cf. bound from Eq.~\eqref{eq:purity_bound}). Inside the unstable region the system size dependence is very weak and we were not able to characterize the asymptotic limit of the purity with the system sizes explored in this work.

\begin{figure}[t]
 \centering
 \includegraphics[width=0.48\textwidth]{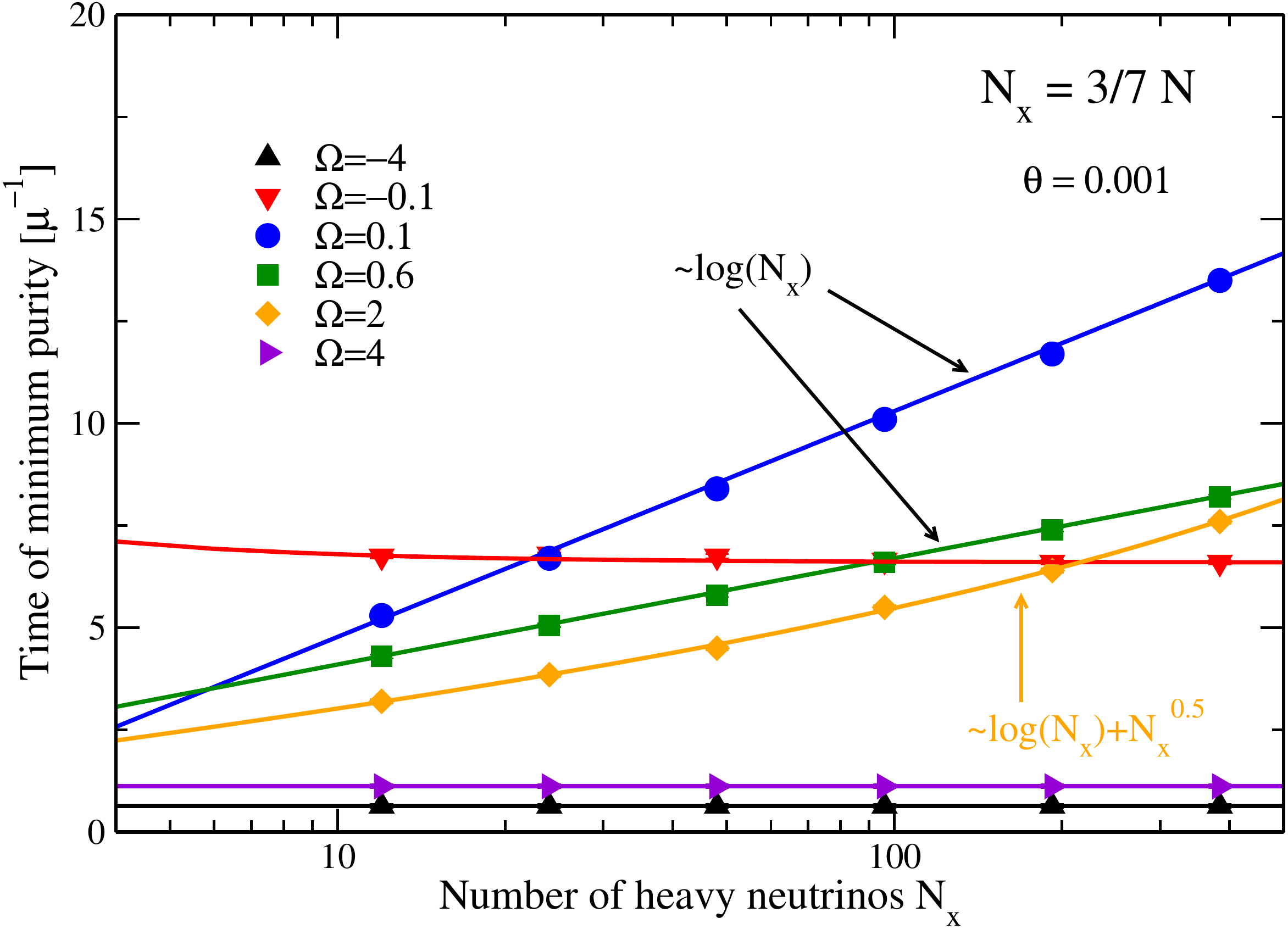}
 \caption{Time to reach minimum of purity for a fixed ratio $N_x=3N/7$ and mixing angle $\theta=0.001$. The unstable region corresponds energy asymmetries $\Omega\in[0.01,1.99]$.}
\label{fig:time_purity_gr_log}
\end{figure}

In panel (b) we show the difference in purity found for a system with $N_e=N_x=1024$ when increasing the value of the mixing angle $\theta$, we can see that the left boundary at $\Omega=0$ remains sharp while there is a smoother transition to the large $\Omega$ regime. Finally, in panel (c) we show the time evolution of the purity for three different values of the energy asymmetry across the transition at $\Omega=0$. For positive values of $\Omega$, inside the unstable region, we find a much quicker development of entanglement than at the critical value $\Omega=0$. For both systems with $\eta=1/2$ as in Fig.~\ref{fig:purity}, as well as other asymmetries, we find that the minimum of the purity is reached on time scales $t_{pmin}\mu\approx\log(N)$ while in the stable region the minimum is attained at constant time. Both of these are similar to the behavior of the entanglement entropy observed in earlier work~\cite{roggero2021entanglement,roggero2021dynamical}. In Fig.~\ref{fig:time_purity_gr_log} the time scale for the minimum is reported for a system near equipartition with $N_x=3/7N$ and for a small mixing angle $\theta=0.001$.

\begin{figure}[t]
 \centering
 \includegraphics[width=0.49\textwidth]{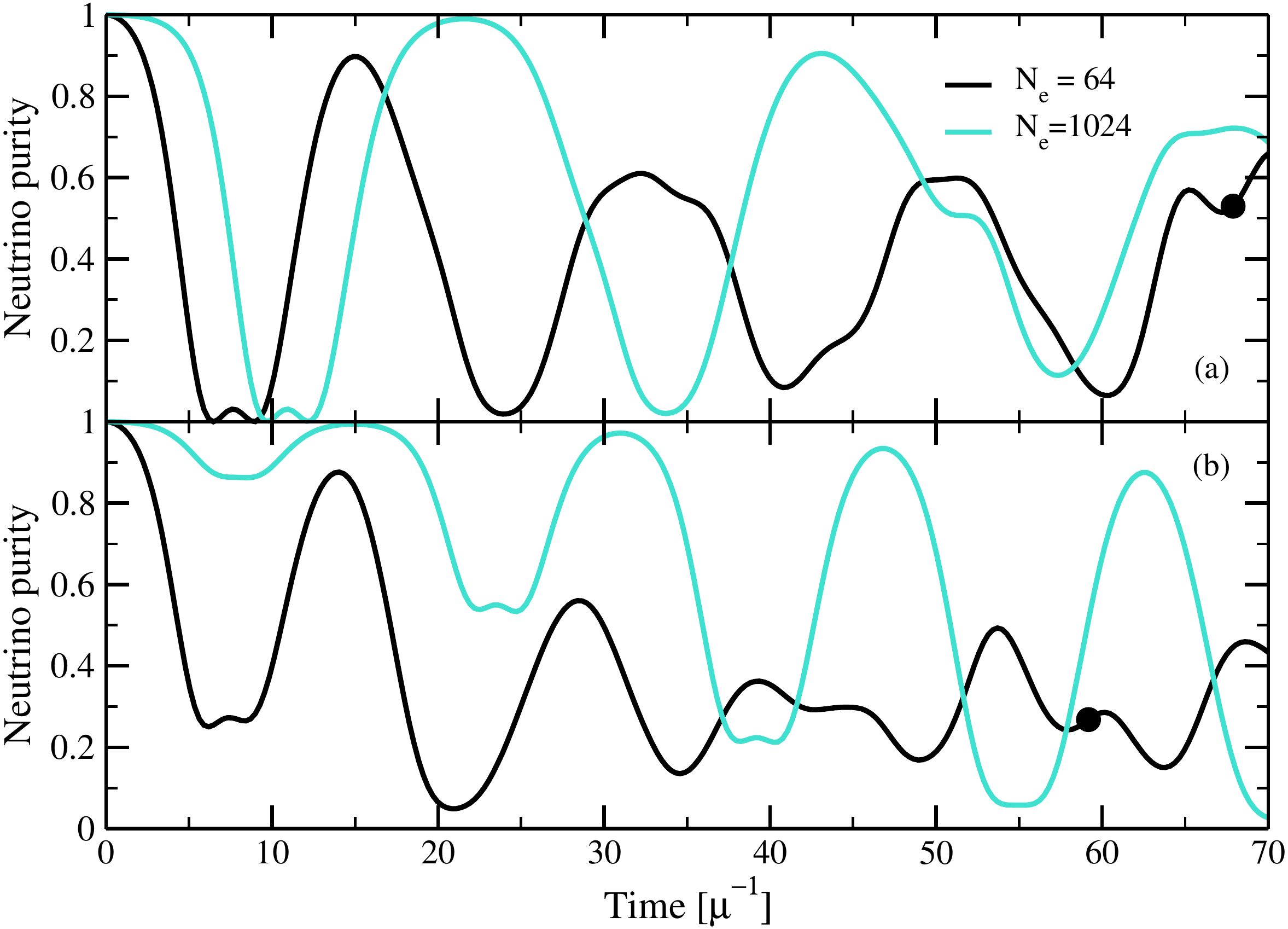}
 \caption{Time evolution of the purity in two models showing bipolar oscillations: panel (a) corresponds to a small mixing angle $\theta=0.001$ and panel (b) to a large mixing angle $\theta=0.1$. The black solid lines correspond to $N_e=N_x=64$ while the turquoise to $N_e=N_x=1024$. The black circles indicate the time scale $t_f$ from Eq.~\eqref{eq:tf} for the models with small system size. The value of $t_f$ for the larger models are out of scale and correspond to $t_f=271.5\mu^{-1}$ for the small mixing angle case and $t_f=110.4\mu^{-1}$ for the larger mixing angle. These two models are the same as those shown in Fig~\ref{fig:xyz_bipolar}.}
\label{fig:bipolar_purity}
\end{figure}

In order to show the dependence of the result with the mixing angle, which can be understood to be a consequence of the scaling of the second moment as discussed in Sec.~\ref{ssec:bipolar_mf}, we present in Fig.~\ref{fig:bipolar_purity} results for the time evolution of the purity in the two bipolar models used in Fig.~\ref{fig:xyz_bipolar} above. Panel (a) shows result for $\theta=0.001$ while panel (b) corresponds to systems with $\theta=0.1$. The system size are the same used in Fig.~\ref{fig:xyz_bipolar} where the solid black line corresponds to $N_e=64$ and the turquoise solid line to $N_e=1024$ and the asymmetry in neutrino populations is $\eta=1/2$. As shown in Fig.~\ref{fig:xyz_bipolar}, the evolution of the three cartesian components of the polarization vector with system size is widely different for these two model: for the small mixing angle the results deviate significantly from the mean-field prediction with these system sizes while for the large mixing angle the simulation with $N_e=1024$ follows closely the mean-field value up to times $t\approx15\mu^{-1}$. As shown in Fig.~\ref{fig:bipolar_purity}, the neutrino purity in both cases shows still significant deviations from the mean-field value $\mathcal{P}=1$ but with important differences depending on the mixing angle.
\begin{figure}[b]
 \centering
 \includegraphics[width=0.48\textwidth]{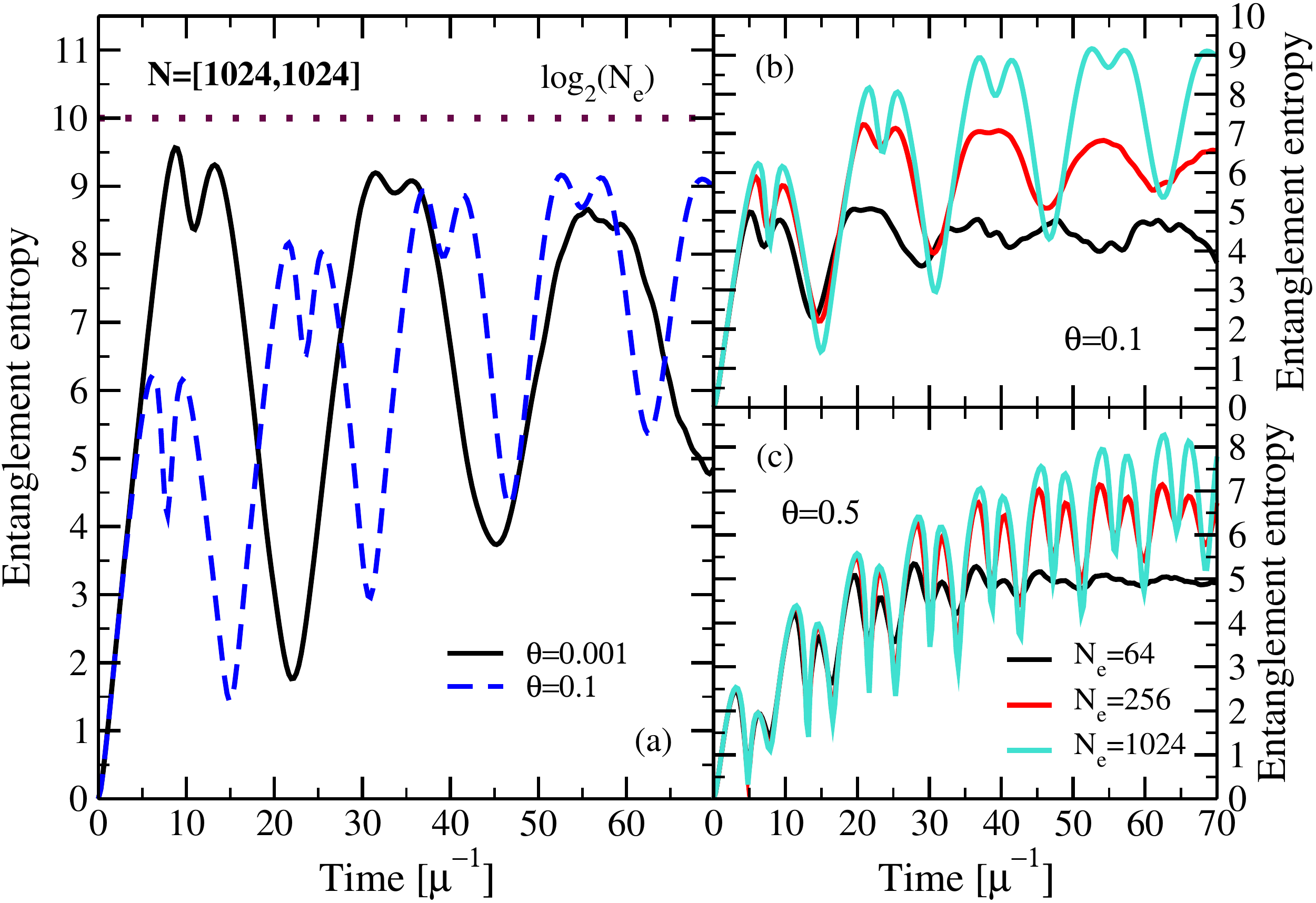}
 \caption{Panel(a) shows the time evolution of the entanglement entropy for the same systems considered in Fig.~\ref{fig:xyz_bipolar} with $N_e=N_x=1024$, $\Omega=0.5$ and $\theta=0.001$ (black solid line) or $\theta=0.1$ (blue dashed line). Panels (b) and (c) show the entropy for 3 different system sizes: $N_e=64$ (black line), $N_e=256$ (red line) and $N_e=1024$ (turquoise line). Panel (b) is for $\theta=0.1$ while panel (c) for a larger value $\theta=0.5$.}
\label{fig:ent_for_fig7}
\end{figure}

In the small mixing angle regime (panel (a)) the minimum of the purity is reached at the first oscillation, its value almost constant and the time scale growing as $\log(N)$ as expected from the previous results shown in Fig.~\ref{fig:time_purity_gr_log}. The behavior of the purity in the large mixing angle case (panel (b)) is instead much different with the first minimum quickly converging to one and little evolution in the time scale. This is seemingly in conflict with the behavior shown in panel (b) of Fig.~\ref{fig:purity} where for $\Omega=0.5$ the purity shows little evolution with the mixing angle. The reason for this is the fact that the purity results in Fig.~\ref{fig:purity} show the minimum reached over a fixed time interval of size $70\mu^{-1}$ and, as we see in panel (b) of Fig.~\ref{fig:bipolar_purity} the minimum is attained at longer times as we increase the system size ($t_{min}\approx20\mu^{-1}$ for $N_e=64$ and $t_{min}\approx70\mu^{-1}$ for $N_e=1024$). This suggests that, for systems showing a quick convergence to the mean field, long time evolution might be needed to use the behavior of the purity to identify unstable conditions. The choice of $70\mu^{-1}$ as the time interval to study flavor evolution was motivated by ensuring that the entanglement entropy could reach its maximum value for the largest system considered here. The time evolution of the entanglement entropy for these two configurations and $N_e=1024$ is shown in panel (a) of Fig.~\ref{fig:ent_for_fig7}, the solid black line corresponds to $\theta=0.001$ while the dashed blue line to $\theta=0.1$. The most significant difference between these results is that the maximum of the entanglement entropy $\approx\log_2(N_e)$ is reached at the first peak for $\theta=0.001$, consistently with what was found at $\theta=0$ in Refs.~\cite{roggero2021entanglement,roggero2021dynamical}, while for larger mixing angles the maximum is reached only after the fourth oscillation. Interestingly, the value of the entropy in the first peaks seems to stop scaling proportionally to $\log(N_e)$ (as the maximum does) and reaches a constant value for large enough system sizes. For small mixing angles instead all peaks seem to scale logarithmically in system size. The convergence of the entropy value as a function of system size is shown more explicitly in panel (b) of Fig.~\ref{fig:ent_for_fig7} where we compare results obtained with $N_e=[64,256,1024]$ and $\theta=0.1$.
\begin{figure}[t]
 \centering
 \includegraphics[width=0.49\textwidth]{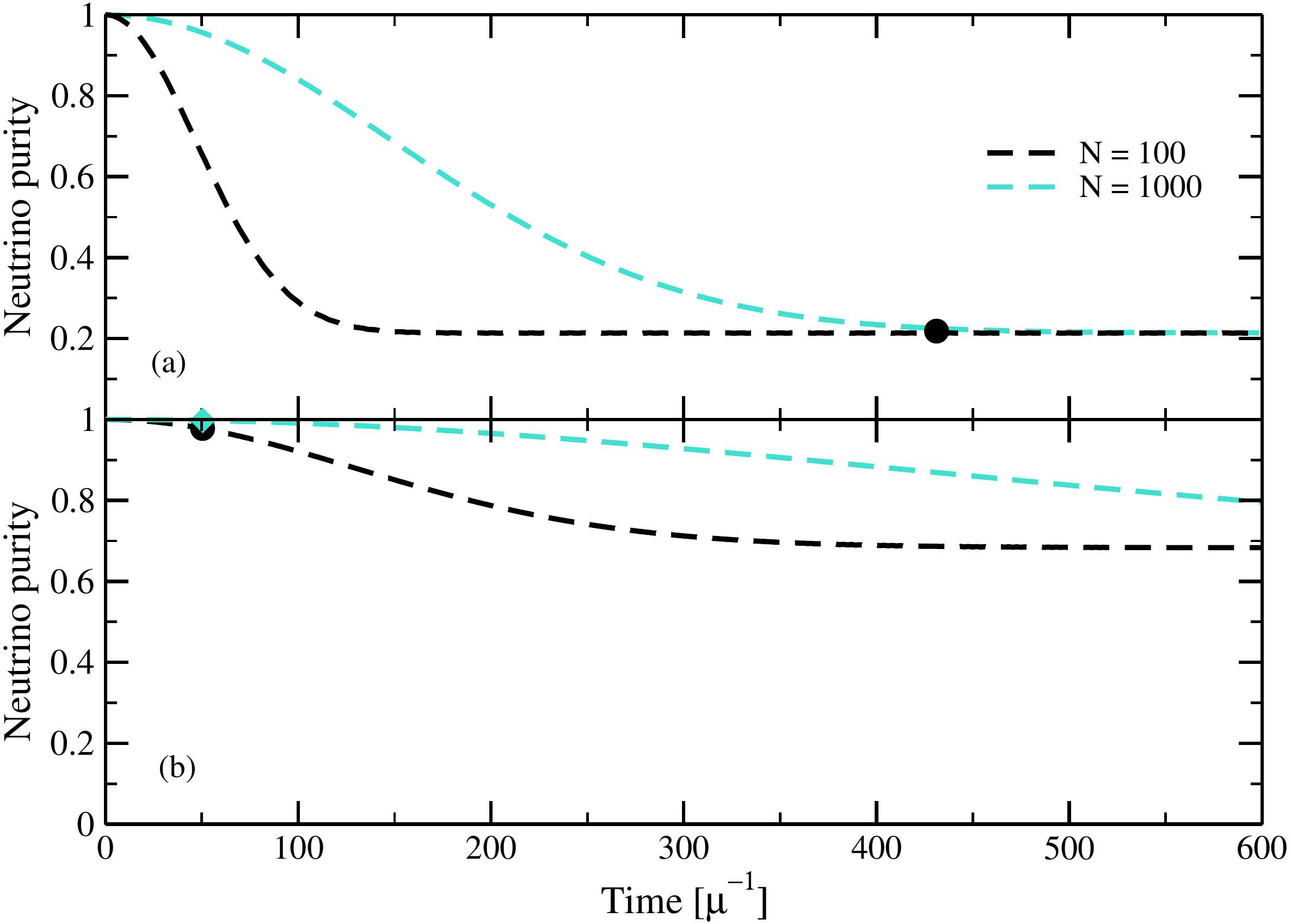}
 \caption{Time evolution of the purity in two models showing precession solutions: panel (a) corresponds to Case 7 in Tab.~\ref{tab:precession} with $\Omega^c\approx1.35\times10^{-4}$ and panel (b) to case 1 with a much larger precession frequency $\Omega^c\approx2.77\times10^{-1}$. The black solid lines correspond to a total system size $N=100$ while the turquoise to $N=1000$. We also indicate with black circles the time $t_f$ from Eq.~\eqref{eq:tf} for the small models with $N=100$ and with a turquoise diamond the $t_f$ time for $N=1000$ in the model of Case 1 (for this model $t_f(N=1000)-t_f(N=100)\approx0.02\mu^{-1}$). For Case 7 we have $t_f=1363\mu^{-1}$ and is out of scale. These two models are the same as those shown in Fig~\ref{fig:xyz_precession}.}
\label{fig:precession_purity}
\end{figure} 

Similarly to the expectation values shown in Fig.~\ref{fig:xyz_bipolar}, we observe good convergence for the first two oscillations while deviations persist at longer times. These results suggest that, in the large system size limit and for finite mixing angles, the entanglement entropy presents oscillations on a time-scale similar to the one for flavor oscillations with maxima which are system size independent but increasing at each oscillation period until eventually reaching the expected value $\approx\log(N_e)$ at long times. In order to test this scenario we have also simulated the entropy evolution for a mixing angle $\theta=0.5$, for which the convergence to the mean-field behavior is much faster, and show the results with different system sizes in panel (c) of Fig.~\ref{fig:ent_for_fig7}. As expected the entropy shows an increase at every oscillation with peaks at late time displaying a slower convergence and reaching values close the maximum. This shows that it is not necessarily correct to understand the results presented in Sec.~\ref{ssec:mf_from_manybody} as a full convergence to the mean field state in the large system size limit since the full evolution creates states with non-zero entanglement. A possibly better characterization is that the mean field predictions of one-body observables become quantitatively correct in the large system size limit, at least for short enough times. For astrophysical neutrinos, of which we do not have direct access to many-body observables, the effect of entanglement might not be observable in practice.

These results presented in this section support the intuition gained in previous work with MPS in Refs.~\cite{roggero2021entanglement,roggero2021dynamical} that entanglement properties in out-of-equilbrium neutrino systems can serve as a diagnostic for the presence of unstable modes. The precession modes may evade this classification as it is currently unclear under what conditions the precession modes are unstable to perturbations. Furthermore, we observe that the presence of collective precession modes is not correlated with maximization of entanglement. However, the time evolution of entanglement can still be useful to uncover characteristic time scales in this regime. In order to explore two extreme regimes, we will now look at entanglement properties of Case 1 (corresponding to $c_0/c_1\approx 0.09$) and of Case 7 (corresponding to $c_0/c_1\approx 1.25 \times 10^{6}$) characterized by the parameters shown in Tab.~\ref{tab:precession} above.

We present in Fig.~\ref{fig:precession_purity} the purity of the majority species for the same simulations used to show the evolution of the flavor polarization in Fig.~\ref{fig:xyz_precession} in Sec.~\ref{ssec:precession_mf}. The time interval $600\mu^{-1}$ is the same used there. An interesting feature that can be noticed from these results is that the typical time scales for entanglement development for both cases is similar while the time scale for flavor oscillations (controlled by the precession frequency $\Omega^c$) are significantly different (cf. results in Fig.~\ref{fig:xyz_precession}). Similarly, the evolution of entanglement also occurs on similar time scales and, for the system sizes explored in this work, saturates over a comparable interval.

\section{Conclusions}
\label{sec:conclusions}

We have studied a simple two-beam model of coherent neutrino oscillations starting
with the dynamics of  an SU(2) Hamiltonian with one- and two-body couplings~(Eq.~\ref{eq:twoBeamH}) with a symmetric initial state within each beam~(Eq.~\ref{eq:genIC}).  The symmetries of this Hamiltonian and initial state
severely limit the propagated amplitudes from  $2^N$ amplitudes to $\mathcal{O}(N^{3/2})$ thus enabling numerical solutions through simple diagonalization of tridiagonal matrices.

We find that the dynamics of one-body observables follow the mean-field (product state) time evolution in the limit of a large number of neutrino flavor spins despite the development of entanglement in the full many-body state. The approach to the mean-field limit with increasing $N$ can depend very sensitively on the initial state. The difference between maximum and minimum energy levels of
the Hamiltonian is proportional to $N$, while the width of an initial product state in the energy space is proportional to $\sqrt{N}$ and the typical level spacing is of order $1/N$. The moments of the energy distribution in the evolving state behave like a Gaussian for moments up to four. When the initial state average energy is in regions where the (symmetric) density of states is low, the convergence to the mean field is comparatively slow.

This Hamiltonian is closely related to typical lattice spin models like the Heisenberg models, though in principle the neutrino Hamiltonian has all-to-all
couplings. The dynamics of these systems may be suitable for studies in trapped-ion or other similar experimental facilities. There are likely physical systems where the finite $N$ results are important, with significant differences
between mean-field (product state) evolution and full quantum simulations. 

We have also studied quantum information measures of the evolving states including entanglement entropy and purity, and find that they can be useful in
identifying regions where collective modes are present. In some cases we find
intriguing relations between the moments of the energy distribution and
simple physical observables. Further study of the relation between quantum information  and physical measures in these systems are warranted.
Generalizations of this problem may be useful for studying both dynamical phase transitions and the eigenstate thermalization hypothesis.
The impact of breaking the symmetries in this system, both in the initial
state and the Hamiltonian, are also very intriguing. In core-collapse supernovae,
these would be introduced by fluctuations in the emission from the proto-neutron star surface and turbulence and time-dependence of the neutrino flux, respectively. 

\begin{acknowledgements}
We thank Ramya Bhaskar, Jeffrey Cohn, Ivan Deutsch, Mario Motta, and Martin Savage for useful discussions and feedback. This work was supported in part by the
Quantum Science Center (QSC), a National Quantum Information Science Research Center of the U.S. Department of Energy (DOE) (J.C and V.C)
and by the U.S. Department of Energy, Office of Science, Office of Nuclear Physics, Inqubator for Quantum Simulation (IQuS) under Award Number DOE (NP) Award DE-SC0020970 (A.R). This work was also supported by the U.S. Department of Energy, Office of Science, Office of Nuclear Physics under Award Number DE-SC0017803 (H.D.) and under Contract Number DE-AC52-06NA25396 (J.D.M).
\end{acknowledgements}

\appendix

\section{Bound on single neutrino purity}
\label{app:purity bound}
We denote by $S_i$ the entropy of the $i$-th neutrino in one of the two beams, let's call it beam $\A$ with $N_\A$ amplitudes and entropy $S_\A$. Using the subadditivity of the entropy we have that
\begin{equation}
S_\A \leq \sum_{i=1}^{N_\A}S_i = N_\A S_1\;,
\end{equation}
where in the second equality we used the permutation symmetry for neutrinos in the beam. This directly implies
\begin{equation}
\label{eq:bound_entropy}
S_1 \geq \frac{S_\A}{N_\A}\;.
\end{equation}
To see how this can be used to place an upperbound on the purity $\mathcal{P}_1$ consider the following parametrization for a single spin density matrix
\begin{equation}
\rho_1 = U_1 \begin{pmatrix}
r&0\\
0&(1-r)\\
\end{pmatrix} U_1^\dagger\,,
\end{equation}
with $r\in[0,1]$ and $U_1$ a unitary transformation to bring a general density matrix to this form. Both the entropy and the purity won't depend on the choice of $U_1$ and we will then neglect it from here on. Also, we can always consider $r\in[0,1/2]$ since we can reparamterize it mapping $r\to1-r$.
We now have
\begin{equation}
S_1 = -r\log_2(r) -(1-r)\log_2(1-r)
\end{equation}
for the entropy and
\begin{equation}
\mathcal{P}_1 = 2\left(r^2 + (1-r)^2\right)-1
\end{equation}
for the purity as defined in Eq.~\eqref{eq:def_purity} of the main text. For $r\in[0,1/2]$ we have
\begin{equation}
S_1 < -r\log_2(r) +r/\ln(2)\;.
\end{equation}
We now express $r$ as follows
\begin{equation}
r=2^{-\alpha}\quad\alpha\geq1\;. 
\end{equation}
The bound on the entropy than becomes
\begin{equation}
S_1<\left(\alpha+\frac{1}{\ln(2)}\right)2^{-\alpha}\;.
\end{equation}
Using the bound in Eq.~\eqref{eq:bound_entropy} we have
\begin{equation}
\left(\alpha+\frac{1}{\ln(2)}\right)2^{-\alpha}>\frac{S_\A}{N_\A}\;.
\end{equation}
Inverting the relationship with $\alpha$ we then find
\begin{equation}
\alpha+\frac{1}{\ln(2)}<-W_{-1}\left(-\frac{S_\A}{eN_\A}\right)\;,
\end{equation}
with $W_{-1}(x)$ the negative branch on the Lambert W function. Using the upperbound~\cite{LambertW2013} for $u>0$
\begin{equation}
1+\sqrt{2u}+u>-W_{-1}\left(-e^{-u-1}\right)\;,
\end{equation}
we find
\begin{equation}
\alpha<1-\frac{1}{\ln(2)}+\sqrt{2\ln\left(\frac{N_\A}{S_\A}\right)}+\ln\left(\frac{N_\A}{S_\A}\right)\;.
\end{equation}
This implies
\begin{equation}
\label{eq:rbound}
r=2^{-\alpha}\geq \frac{e}{2}\left(\frac{S_\A}{N_\A}\right)^{\ln(2)}2^{-\sqrt{2\ln\left(\frac{N_\A}{S_\A}\right)}}\;.
\end{equation}
We are now in a position to place an upperbound on the purity. Since, for $r\in[0,1/2]$ the purity is monotonically decreasing in $r$ we can directly use the lower bound in Eq.~\eqref{eq:rbound}. A more manageable expression, which is however not very tight, is to use instead
\begin{equation}
r^2 + (1-r)^2<1-r\;,
\end{equation}
to arrive at
\begin{equation}
\mathcal{P}_i < 1-e\left(\frac{S_\A}{N_\A}\right)^{\ln(2)}2^{-\sqrt{2\ln\left(\frac{N_\A}{S_\A}\right)}}\;.
\end{equation}
This is the bound quoted in the main text.

\bibliography{refs.bib}

\end{document}